\documentclass[journal]{vgtc}                     

\onlineid{0}

\graphicspath{{figs/}{figures/}{pictures/}{images/}{./}} 

\usepackage{tabu}                      
\usepackage{booktabs}                  
\usepackage{lipsum}                    
\usepackage{mwe}                       
\usepackage{paralist}

\usepackage{mathptmx}                  
\usepackage{comment}

\usepackage{microtype}                 
\PassOptionsToPackage{warn}{textcomp}  
\usepackage{textcomp}                  
\usepackage{mathptmx}                  
\usepackage{times}                     
\usepackage{cite}                      
\usepackage{tabu}                      
\usepackage{booktabs}                  
\usepackage{algorithm}
\usepackage{algpseudocode}             
\usepackage{wrapfig}
\newcommand{\systemname}{{ActorLens}}
\newcommand{\name}{{\textit{ActorLens}}}
\newcommand{\vone}{Projection View}
\newcommand{\vtwo}{Progression View}
\newcommand{\vthree}{Summary View}
\newcommand{\vfour}{Match Replay View}

\newcommand{\appendixdata}{Appendix A} 
\newcommand{\appendixevents}{Appendix B}
\newcommand{\appendixcasetwo}{Appendix C}
\newcommand{\appendixlowlevel}{Appendix D}
\newcommand{\appendixxgboost}{Appendix E}

\AtBeginDocument{

}

\newcommand{\mzhihua}[1]{#1}

\newcommand{\dq}[1]{``#1''}

\vgtccategory{Research}

\vgtcpapertype{Application}

\title{{\systemname}: Visual Analytics for High-level Actor Identification in MOBA Games}

\author{%
  Zhihua Jin\thanks{Co-first authors.}, Gaoping Huang\footnotemark[1], Zixin Chen, Shiyi Liu, Yang Chao\thanks{Co-corresponding authors.}, Zhenchuan Yang, Quan Li\footnotemark[2], and Huamin Qu
}
\authorfooter{
  \item
  	Zhihua Jin, Zixin Chen, and Huamin Qu are with Hong Kong University of Science and Technology.
   \protect\\
  	E-mail: \{zjinak, zchendf, huamin\}@cse.ust.hk.
  \item
  	Gaoping Huang, Yang Chao, and Zhenchuan Yang are with Tencent.
   \protect\\
  	E-mail: \{gavinphuang, youngchao, zachyang\}@tencent.com.

  \item Shiyi Liu and Quan Li are with ShanghaiTech University.
  \protect\\
  	E-mail: \{liushy, liquan\}@shanghaitech.edu.cn.
}

\abstract{Multiplayer Online Battle Arenas (MOBAs) have garnered a substantial player base worldwide. Nevertheless, the presence of noxious players, commonly referred to as ``actors'', can significantly compromise game fairness by exhibiting negative behaviors that diminish their team's competitive edge. Furthermore, high-level actors tend to engage in more egregious conduct to evade detection, thereby causing harm to the game community and necessitating their identification. To tackle this urgent concern, a partnership was formed with a team of game specialists from a prominent company to facilitate the identification and labeling of high-level actors in MOBA games. We first characterize the problem and abstract data and events from the game scene to formulate design requirements. Subsequently, \textit{{\systemname}}, a visual analytics system, was developed to exclude low-level actors, detect potential high-level actors, and assist users in labeling players. \textit{{\systemname}} furnishes an overview of players' status, summarizes behavioral patterns across three player cohorts (namely, focused players, historical matches of focused players, and matches of other players who played the same hero), and synthesizes key match events. By incorporating multiple views of information, users can proficiently recognize and label high-level actors in MOBA games. 
We conducted case studies and user studies to demonstrate the efficacy of the system.
}


\teaser{
  \centering
  \includegraphics[width=\linewidth]{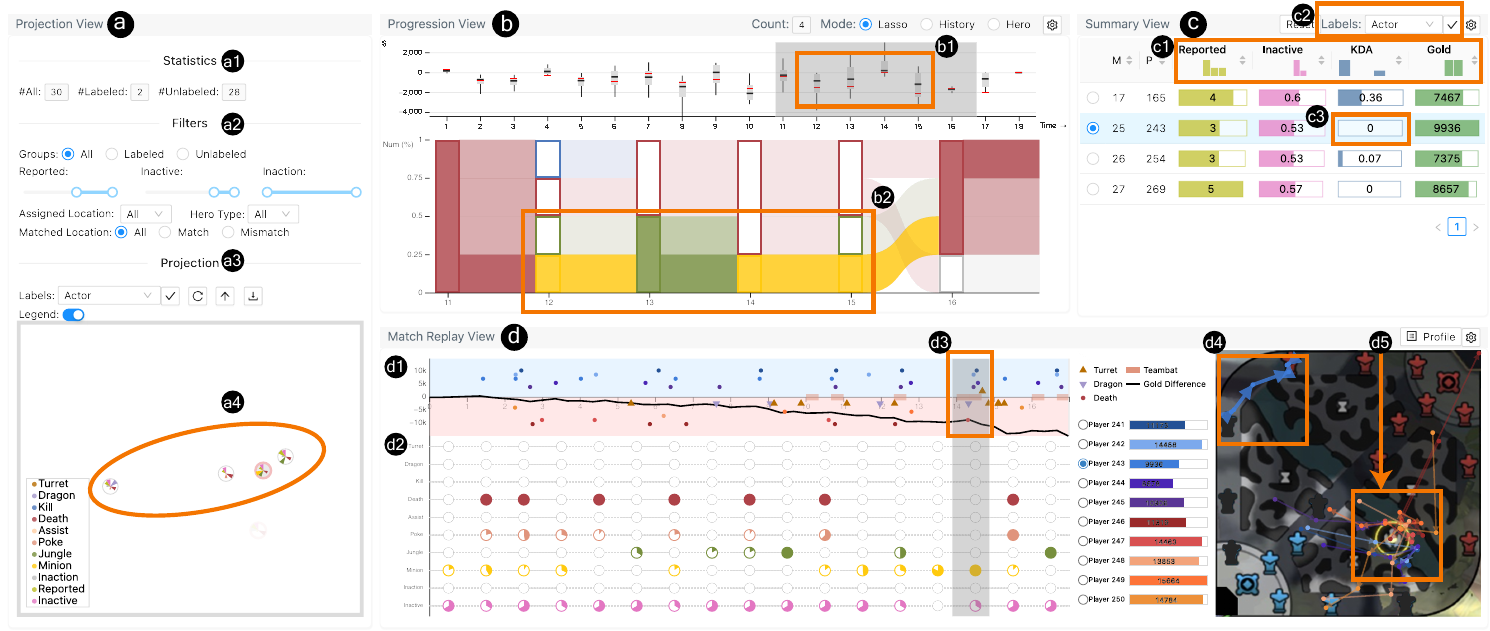}
  \vspace{-6mm}
  \caption{%
  	{\name} is designed to help users identify and label high-level actors in MOBA games. It consists of {\vone}, {\vtwo}, {\vthree}, and {\vfour}. (a) {\vone} provides an overview of the status of focused players across the matches. (b) {\vtwo} summarizes the behaviors of players in multiple matches. (c) {\vthree} provides summary information for focused players and helps users narrow down. (d) {\vfour} reproduces the whole battle progress and enables users to check the detailed events of one player and contextualizes its trajectory with other players.
  }
  \label{fig:teaser}
}

\begin{document}




\firstsection{Introduction}
\maketitle

\par Multiplayer Online Battle Arenas (MOBAs) represent a popular genre of esports. Notably, several prominent games within this category, including \textit{Defense of the Ancients 2 (DotA2)}, \textit{League of Legends (LoL)}, and \textit{Honor of King (HoK)}, have garnered millions of viewers through hosting professional tournaments on an international scale~\cite{Kokkinakis2020-uo}. A standard match in a MOBA game consists of two teams comprised of ten players, with each player controlling a distinct character, also known as a \textit{hero}. In order to win, players must work together to attack the opposing team and destroy their crystal. Success in a MOBA game relies heavily on strategic gameplay tactics and cooperative teamwork~\cite{Martens2015-qj}.

\par However, it has been observed that some individuals may engage in malicious behavior that disrupts the cooperative nature of the game and influences the outcome through improper means. For instance, players who do not share the same goals as the rest of the team may act contrary to the team's overall objectives~\cite{kwak2015exploring,Kou2020-eh}, ultimately leading to unfairness and emotional distress for other players. One such behavior involves players deliberately staying away from their keyboard (AFK) by staying in the highland during gameplay, which can significantly impede the progress of the team and disrupt the overall player experience. The existence of these toxic behaviors undermines the fairness of the match and can result in detrimental consequences for players, including loss of the game~\cite{kwak2015exploring}. These toxic players are formally defined as \textbf{actors}, and they engage in disruptive behaviors that undermine the integrity of the game. Specifically, \textbf{low-level actors} typically exhibit one of two behaviors: going AFK or intentionally rushing toward the opposing team's tower in order to be killed (known as feeding). In contrast, \textbf{high-level actors} exhibit more subtle disruptive behaviors that are not as readily apparent as those exhibited by low-level actors. For instance, the in-game statistics of high-level actors, particularly the combination of kills, deaths, and assists (KDA), may appear no different from those of regular players. However, high-level actors deliberately avoid participating in crucial events of the game, such as team fights, ultimately leading to a loss for their team or a negative experience for their fellow teammates. Detecting high-level actors can be particularly challenging, as they possess a wealth of gameplay experience and expertise that enables them to evade detection through simple rules and measures. Ultimately, the cheating behaviors of high-level actors blur the boundaries of their definition as skilled players and pose a significant challenge to the maintenance of fairness and integrity in online MOBA gaming.

\par In order to gain insight into the challenges associated with identifying higher-level actors in online gaming environments, we conducted a series of in-depth interviews with domain experts at a prominent gaming company. Through these discussions, we discovered that a variety of algorithms~\cite {Canossa2021-vu,Kou2020-eh,Martens2015-qj} and models~\cite{Wong2022-im,Pimentel2020-eh,Geiger2020-ed} exist that are capable of detecting low-level actors. However, these approaches are not necessarily appropriate for detecting high-level actors due to several key factors. First, algorithms designed for automated anomaly detection, such as those used in domains like environment~\cite{Russo2020-xi} and city planning~\cite{Lin2018-er}, typically require explicit input features and labels, which can be ambiguous or difficult to define in the context of high-level game actors. Second, automated solutions may not provide a clear interpretation of their predictions, which limits their usefulness in complex environments like MOBA games. Currently, high-level actors in MOBA games are typically identified through a manual reporting and review process. This approach involves players reporting suspicious behaviors, which are then reviewed by game inspectors who analyze the game record to make a judgment. However, this process is often time-consuming and subjective, as it relies heavily on the experience and judgment of the reviewer. In some cases, an unreasonable or problematic review can even worsen the game experience for both the reporter and the reported player, leading to further consequences such as player loss~\cite{pohjanen2018report}. Given these challenges, it is clear that a more efficient, objective, and evidence-based review solution is needed to accurately identify high-level actors in online gaming environments.

\par To identify high-level actors and establish their behavioral boundaries, we must first address the following three challenges: 1) \textbf{Ambiguity in the feature space}, resulting from the imprecise and challenging-to-translate descriptions of high-level actors provided by domain experts. For instance, a high-level actor may appear to perform well with a high damage indicator, despite intentionally attacking only unimportant opposing heroes. Algorithms that rely solely on such indicators are thus incapable of accurately identifying high-level actors. Therefore, it is crucial to explicitly define and characterize high-level actors. 2) \textbf{Inefficiency in actor identification}, largely due to the reliance on human experience and the time-consuming nature of existing solutions. A game inspector must review the entire game replay, which takes about $20$ minutes on average, making it a tedious task. 3) \textbf{The need for comprehensive quantitative data or evidence to support actor identification.} While general algorithms may focus on detecting deviations in player behavior compared to other players in the current state~\cite {Canossa2021-vu,Kou2020-eh,Martens2015-qj}, interactive anomaly detection solutions are typically designed to aid automated algorithms in determining anomalies through human experience and improving accuracy. However, most such solutions focus solely on distinguishing between a single individual and other individuals in the current session. To determine whether a player has made a mistake involuntarily or is a high-level actor with a proven record, it is necessary to track the player over a longer period of time and with greater granularity. This is a non-trivial task requiring both longitudinal analyses (comparing players with their own historical data) and cross-sectional analyses (comparing players with other players on the team).

\par In this study, a visual analytics system named {\name} was developed to aid users in identifying and labeling high-level actors in MOBA games (\autoref{fig:teaser}). The problem was defined, and previous workflows for identifying high-level actors were summarized with the help of a team of experts from a leading game company. Design requirements were derived through expert interviews, and match data was collected from the game \textit{League of Legends: Wild Rift}. Low-level actor detection algorithms were designed to filter out low-level actors, and metrics were derived to locate potential high-level actors. Interactive models were also designed to guide users in labeling unlabeled players. The system provides an overview of focused players' status across matches and allows users to label actors with interactive models' guidance. It also summarizes the behavioral patterns of selected players across multiple matches and enables users to inspect three cohorts of players, including selected players from users, historical matches of the selected player, and matches of other players who played the same hero. The system further supports summarizing key events for one player and the trajectories of ten players to help users understand the dynamics of battle progress. Users can combine information from multiple perspectives to determine whether a focused player is a high-level actor and provide a label for them. 
The system's effectiveness and usability were evaluated through case studies and user studies. The contributions of this study are summarized as follows:
\begin{compactitem}
    \item A problem characterization of the task of identifying and labeling high-level actors in MOBA games.
    \item The development of a visual analytics system designed to efficiently perform the task of identifying and labeling high-level actors.
    \item The evaluation of the system's effectiveness and usability through case studies and user studies.
\end{compactitem}

\section{Related Work}
\par Literature that overlaps with this work can be categorized into three groups, namely \textit{anomaly detection and visualization}, \textit{data labeling tools}, and \textit{gameplay data visualization}.

\subsection{Anomaly Detection and Visualization}
\par Anomaly detection, which aims to identify unexpected patterns, has been extensively studied in the past decades. Approaches for anomaly detection can be broadly classified into statistical-based methods~\cite{Rousseeuw2005-hh}, distance-based methods~\cite{Breunig2000-fl,Bay2003-hu}, density-based methods~\cite{Breunig2000-fl,Bay2003-hu}, and cluster-based methods~\cite{Smiti2013-bb,Smiti2012-cf}. Visual analytics provides an opportunity for analysts to be involved in the inspection process and make better decisions. More recent studies proposed systems for analyzing multivariate data~\cite{Liu2021-nb,Zhao2019-rl,Xu2018-ll}, sequence data~\cite{Guo2019-pv,Mu2019-yq}, temporal data~\cite{Xu2020-mu,Cao2016-iq}, and spatiotemporal data~\cite{Cao2017-ki,Chae2012-wa,McKenna2016-py}. These visualization systems have been applied in numerous fields such as urban planning~\cite{Cao2017-ki,Liao2010-oo}, web environment~\cite{McKenna2016-py,Xu2020-mu}, social science~\cite{Chae2012-wa,Zhao2014-xv}, and education~\cite{Mu2019-yq}. For instance, Cao et al.\cite{Cao2017-ki} developed \textit{Voila}, a visualization tool based on taxi data that shows potential anomaly incidents in corresponding abnormal regions to avoid undesired outcomes. Zhao et al.\cite{Zhao2014-xv} introduced \textit{FluxFlow}, a system that integrates machine learning algorithms to detect anomalous information spreading in social media. Nonetheless, previous works did not consider the multivariate and spatiotemporal aspects of game data and did not focus on detecting anomalous behavior in MOBA games.

\par In gaming scenarios, behaviors that deviate from the norm are commonly referred to as toxicity, encompassing deviant behavior~\cite{Grandprey-Shores2014-ra,Kou2017-nx,Foo2004-oi} and abusive language~\cite{Cheng2019-sm,Vaz2021-ka,Martens2015-qj}. Existing research primarily focuses on analyzing the effects~\cite{Zsila2022-ov,Cheng2019-sm}, underlying reasons~\cite{Kordyaka2022-zi}, and coping strategies~\cite{Adinolf2018-wf} for toxicity. Machine learning algorithms have been applied to detect abusive language with high accuracy rates~\cite{Vaz2021-ka,Martens2015-qj}, such as Vaz et al.'s~\cite{Vaz2021-ka} use of KNIME to achieve 92\% and 85\% accuracy for toxic and non-toxic messages, respectively. However, there are limited automated methods for detecting other forms of toxic behavior that manifest through player actions in gameplay. Furthermore, automated anomaly detection via machine learning is often impeded by unclear boundaries and insufficient labeled data~\cite{Hodge2004-ta,Fanaee-T2016-sy,Chandola2009-wv}.

\par This study examines toxic behavior in MOBA games through the use of visual analytics. Specifically, we propose novel interactive designs for detecting anomalies in multivariate spatiotemporal in-game data. The proposed system is intended to aid game designers, inspectors, and algorithm developers in identifying, investigating, and labeling players who exhibit toxic behaviors in MOBA games. By providing an interactive visualization of in-game data, our approach enables users to explore the relationships between different variables and identify anomalous behaviors that may have a detrimental impact on the game experience. Additionally, our system facilitates the tracking of the temporal evolution of toxic behaviors and allows for comparison across different player groups. Our findings suggest that our approach can contribute to the enhancement of the game community's well-being and improve the game experience for all players.

\subsection{Data Labeling Tools}
\par The demand for data labeling tools has been increasing, with many tools being developed to facilitate the labeling process. These tools are typically designed to label various types of data, such as images~\cite{torralba2010labelme,deng2014scalable}, audio~\cite{bryan2014isse}, videos~\cite{liao2016visualization,dutta2019via}, and texts~\cite{kucher2017active}. To accelerate the labeling process, techniques like clustering~\cite{suh2007semi,cui2007easyalbum,tang2013towards}, active learning~\cite{settles2009active,liao2016visualization}, and other semi-automatic methods~\cite{paiva2014approach,zhang2021mi3} have been utilized. Furthermore, efforts have been made to alleviate the workload associated with developing data labeling tools. One example is OneLabeler~\cite{zhang2022onelabeler}, which introduced a workflow and tool for assisting users in quickly building data labeling tools. By interactively controlling the components in the panel, OneLabeler can address simple labeling tasks. However, due to the complexity of game data and the associated data processing, these tools may not be suitable for such scenarios.

\par Our work allows users to refine groups of potential high-level actors through an iterative process of interactive visual exploration and labeling. Through the use of a visual analytics solution, our approach enhances the efficiency of the labeling process and provides domain experts with insights into the characteristics and behaviors of high-level actors in the game.

\subsection{Gameplay Data Visualization}

\par The visualization of game data is a common technique employed in game analytics. Wallner and Kriglstein~\cite{Wallner2013-um} provided a comprehensive overview of game data visualizations, categorizing them into various encoding methods such as charts and diagrams~\cite{Milam2010-dw,Medler2011-fd}, heatmaps~\cite{Ashton2011-uc,Drachen2009-rq}, movement visualizations~\cite{Drachen2009-ys}, self-organizing maps~\cite{Thawonmas2006-rd,Drachen2009-bc}, and node-link representations~\cite{Thawonmas2008-ny,Wallner2011-fe}. Past efforts on post-game data visualization were typically focused on player training, but identifying high-level actors necessitates evaluation based on their behavioral performance, which cannot be readily accomplished using post-game data. On the other hand, previous work on in-game data visualization was generally intended to aid game developers in improving user experience or enhancing game mechanics~\cite{Li2017-dn,Javvaji2020-eu,Xie2022-ku,Lu2019-zf,Ahmad2019-uq}. For example, Li et al.\cite{Li2017-dn,Li2018-kb} utilized visualization techniques to comprehend the causes of snowballing and comebacks. This exploration facilitates game developers to refine game balance effectively. For game designers, Javvaji et al.\cite{Javvaji2020-eu} presented an interactive visualization of game telemetry data to learn player patterns. On the other hand, Xie et al.~\cite{Xie2022-ku} introduced \textit{RoleSeer} to investigate informal roles in MMORPGs from the perspectives of behavioral interactions and depict players' dynamic interconversions and transitions. These systems aim to analyze the overall game state but do not permit the exploration of individual player behavior.

\par The availability of large-scale game data and players' growing interest in data have spurred the development of training visualizations, which have received considerable attention~\cite{Wallner2021-fy,Afonso2019-of,Charleer2018-kv,Kuan2017-ye}. Wallner et al.~\cite{Wallner2021-fy} created a training visualization tool that enables players to review their gameplay, allowing them to learn from their experiences. Afonso et al.~\cite{Afonso2019-of} used visual analytics to analyze spatiotemporal in-game data and understand how players deal with game information, with the aim of improving players' strategies. While these approaches examine players' behaviors and strategies, they do not filter or identify abnormal behaviors among them.

\par This study introduces a novel visualization method aimed at assisting game developers or inspectors in analyzing and labeling high-level actors in MOBA games. The proposed system is capable of summarizing multiple matches and inspecting detailed in-game data, thus enabling users to easily capture actor characteristics and label them.

\section{Background}

\par In this study, we first present the mechanisms employed in the MOBA game that we investigate, namely, ``League of Legends: Wild Rift.'' Following this, we delve into the definitions of actors.


\subsection{League of Legends: Wild Rift}
\par \textit{League of Legends: Wild Rift} is a mobile MOBA game\footnote{https://lolm.qq.com/main.html}, wherein two teams, each consisting of five players, engage in gameplay that spans an average duration of 20 minutes. The objective of the game is for one team to emerge victorious by obliterating the adversary's base.

\begin{figure}[h]
  \centering
  \vspace{-3mm}
  \includegraphics[width=\linewidth]{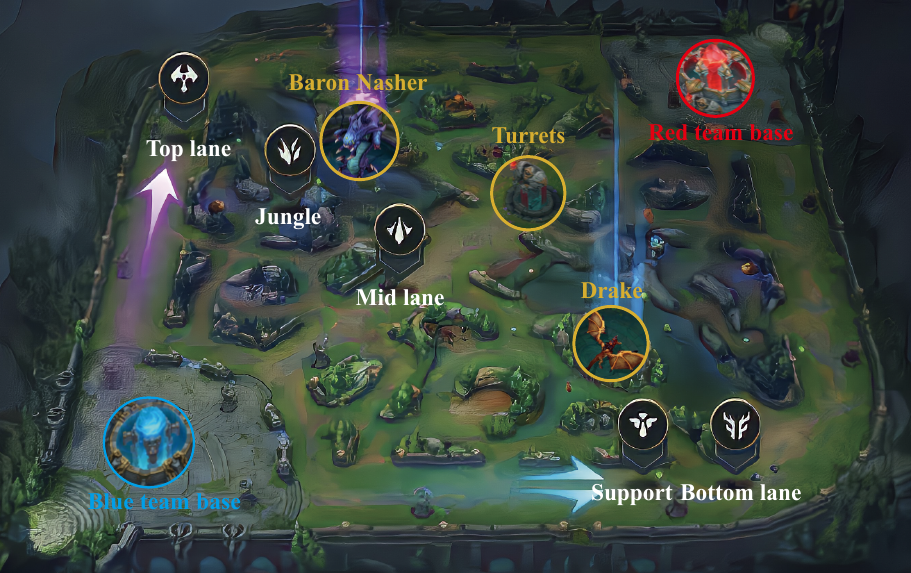}
  \vspace{-6mm}
  \caption{The virtual scene in \textit{League of Legends: Wild Rift}.}
  \label{fig:bg}
    \vspace{-3mm}
\end{figure}

\par Within the virtual environment (\autoref{fig:bg}), two factions are distinguished as the blue team and the red team, each of which commences gameplay from opposite ends of the map. The bases of the respective teams incessantly spawn feeble, automated units knowns as minions, which advance toward the opposing base via three lanes: the top lane, mid lane, and bottom lane. The players endeavor to guide these minions into the adversary's base in order to cause damage to it and, ultimately, to claim victory. Additionally, every lane features three turrets, which initially focus their fire on nearby enemy minions but immediately shift their attention to enemy players if alliance players are attacked. The space lying between the three lanes is composed of neutral regions known as ``jungle areas'', distributed throughout the four quadrants of the map. These jungle areas contain neutral resources, and players can slay monsters to acquire gold. Additionally, there are four monsters, two blue-buff monsters and two red-buff monsters, which confer beneficial effects upon the players who vanquish them. The most potent neutral resources are the drakes and baron nasher, which grant significant positive effects to the entire team upon their defeat.

\par A team may employ various tactics to attain an advantageous position in a competitive game. During the early stages, team members may venture to their respective lanes or jungle regions to accrue monetary rewards by eliminating minions or monsters. When the opportune moment arises, the team can coordinate a team battle to eliminate the opposing heroes, thus enabling them to gain an advantage in both gold and experience. Subsequently, the team can leverage their advantages to more easily eliminate the drake or baron nasher, or decimate the opposing team's turrets. Conversely, if the team finds itself in a disadvantaged situation, they may resort to defensive strategies by safeguarding their turrets and seizing opportunities to counterattack to exploit mistakes made by their adversaries. In the later stages of the game, the team may utilize a single hero to lead the charge and stealthily destroy the opponent's turrets. Through these strategic changes and adjustments, one game will ultimately emerge victorious by obliterating the base of their opponents.




\subsection{Actors Definition}

\par In the context of gaming, actors are individuals who engage in manipulative tactics with the aim of deliberately sabotaging their team's chances of winning. Such individuals may display uncooperative or even supportive behaviors toward their opponents. A common example of this behavior is known as ``feed'', whereby the actor is intentionally killed by enemy heroes without attacking enemy heroes, thereby granting the opponent economic, empirical, and map advantages. The consequences of these behaviors are detrimental to the actor's team, as they often result in the loss of defensive turrets due to the lack of sufficient teammates to defend. Over time, these disadvantages accumulate, making it increasingly challenging for the actor's team to turn the game around in their favor.

\section{Abstractions}
\label{sec:abstractions}


\par During the course of six months, our team engaged in frequent biweekly meetings with three domain experts from a prominent game company that operates the \textit{League of Legends: Wild Rift}. The first expert (\textbf{E1}) was responsible for constructing supportive tools that aid in the analysis of game data, while the second expert (\textbf{E2}) focused on the development of various algorithms aimed at identifying behaviors present in the game. The third expert (\textbf{E3}) served as the team leader and had experience in identifying toxic players across multiple types of games. These collaborative efforts enabled us to comprehensively investigate high-level actors within the domain problem while also identifying key design requirements for our approach.

\subsection{Experts' Conventional Practices and Bottlenecks}

\par In 2021, the release of \textit{League of Legends: Wild Rift} attracted numerous players. Regrettably, the game company has received a plethora of reports from normal players whose experiences have been negatively impacted by certain player behaviors. These players, commonly referred to as ``actors'', deliberately manipulate the outcome of matches in their favor. Consequently, the game company acknowledged these issues and commenced the development of algorithms aimed at detecting and combating them. While some overt behaviors, such as AFK and Feeder, are easily detectable and have been incorporated into the system to prevent these players from continuing to play the game, other types of covert behaviors have emerged to evade these algorithms, ultimately resulting in a negative gaming experience. Players who exhibit such behaviors are classified as high-level actors. Thus, there is an urgent need for the game company to undertake a further in-depth investigation and characterization of high-level actors.


\par To address the aforementioned issues, the game company employed a specific workflow to target high-level actors. First, after filtering out all low-level actors, the annotators carefully reviewed  the entire game video to determine whether or not any actors were present and then composed a brief description for each one. Second, following the annotation process, game designers and algorithm developers convened to identify any new abnormal behaviors, i.e., high-level actor behaviors. Third, each high-level actor's behavior was cross-referenced with multiple matches with corresponding descriptions. Based on these descriptions, algorithm developers endeavored to convert them into specific rule-based algorithms capable of detecting actors. In instances where the descriptions were clear, they can easily readily be converted into rules, and suitable thresholds could be established for the relevant data field to facilitate the implementation of the rules. If, however, the descriptions were vague, algorithm developers conducted multiple rounds of trial and error to refine the rules. It is conceivable that despite following this workflow, suitable rules for detecting certain behaviors remained elusive.

 \par In such a workflow, the game team encountered two challenges. First, the process of scrutinizing lengthy game videos proves to be arduous and time-consuming. 
 The average duration of a game is roughly 20 minutes, which renders complete viewing an inefficient use of time.
 Additionally, increased playback speed may lead to overlooking pivotal moments, resulting in potentially unreliable labeling outcomes. Second, even after discussing the definitions of high-level actor behaviors, converting these descriptions into specific rule-based algorithms remains a formidable task. Due to a lack of robust association with relevant data fields, algorithm developers encounter difficulties in detecting high-level actors. Consequently, they sought a more efficient and intuitive solution to facilitate efficient labeling of high-level actors and permit an easy investigation of their characteristics and boundaries.

\subsection{Experts' Needs and Expectations}

\par In response to the aforementioned challenges in the previous workflow, we distill a set of requirements through interviews with domain experts. The objective was to design a system that would facilitate the identification of potential high-level actors and enable the extraction of their distinguishing features for further examination.
\par \textbf{R1: Locate potential high-level actors after filtering out low-level actors. } Domain experts have highlighted the necessity for a system that is capable of effectively filtering out low-level actors and locating potential high-level actors in a game. 
In order to fulfill this objective, the system must undertake an initial analysis of game data to identify players that exhibit high-level behaviors. Moreover, to enable further exploration, the system should present a user-friendly method for locating potential high-level actors. Specifically, we need to answer the following two questions:
\begin{compactitem}
\item \textit{\textbf{Q1.1}: Which players remain after filtering out multiple conditions?} In order to filter potential high-level actors in a game, various conditions can be utilized, such as the number of reports received by a player, their inactivity score, or their selected location. By applying these conditions, users can effectively narrow down the search to specific groups of players for an initial inspection.
\item \textit{\textbf{Q1.2}: How to group players with similar behaviors together?} To facilitate efficient inspection of behavior patterns, the system should enable users to group players with similar behaviors together. This would assist in identifying commonalities among players and provide a more comprehensive understanding of their gameplay strategies.
\end{compactitem}

\par \textbf{R2: Summarize players' behaviors in multiple matches.} Upon selecting a target cohort of players, it is imperative for experts to obtain a synopsis detailing the alterations in the behaviors of said players throughout a range of matches. Such summaries must effectively elucidate the actions executed at particular time intervals and the impact thereof on the team's competitive edge. By means of these condensed data, users are able to discern exceptional occurrences exhibited by potential high-level actors and subsequently pinpoint the precise timing of such behaviors.
\begin{compactitem}
\item \textit{\textbf{Q2.1}: What are the prevailing patterns of behavior exhibited within a cohort of players?} 
The experts exhibit keenness toward analyzing the collective conduct of a targeted cluster of players across several matches. Users are able to specify the aforementioned cluster of players, and subsequently, concise and informative data summaries must be furnished to facilitate the study of the high-level actors.
\item \textit{\textbf{Q2.2}: What are the prevalent patterns of conduct exhibited by a single player across historical matches?} Given the typical occurrence of multiple matches played by players in the preceding week, a synopsis is required to scrutinize the alternations in the players' behavioral tendencies over the same period. This examination enables domain experts to identify any atypical occurrences or deviations in the player's past performance.
\item \textit{\textbf{Q2.3}: What are the prevalent patterns of conduct observed in the historical matches of a given hero?} With the hero's strategy likely exhibiting greater consistency across multiple typical players, the system ought to furnish a comprehensive overview of the conventional tactics employed in playing the hero, thereby facilitating an investigation into the customary behavior of players. By comparing the established strategy with the actions of a specific player, domain experts can detect any deviant actions.
\end{compactitem}

\textbf{R3: Ascertain the specific behaviors exhibited by individual players during a single match.} To achieve this, experts compare the tactics utilized by the focal player with those commonly used by the group of players under investigation, the past performances of individual players, and the typical strategies employed for a given hero. With this initial analysis, the experts can identify actions that deviate from established norms and require more detailed scrutiny. The system must provide a comprehensive overview of the focal player's performance in the particular match, including specific events and patterns of movement. By contrasting the tactics employed by the focal player with those used by other players in the same match, the experts can gain an in-depth understanding of the player's intention for each action.  Based on this analysis, experts can identify abnormal events or behaviors that may signify a player is a high-level actor.

\begin{compactitem}
\item \textit{\textbf{Q3.1}: How do player actions evolve over time in a match?} Given that players can undertake numerous actions within a single match, lasting approximately 20 minutes on average, the nature of their actions may alter over time. High-level actors may obfuscate their negative conduct as commonplace, therefore, an in-depth comprehension of the dynamic nature of player actions over the course of a match is necessary for users to make appropriate evaluations.
\item \textit{\textbf{Q3.2}: How do team members influence a player's decision-making during a match?} As the battle progresses, the team's performance may fluctuate, potentially altering the course of the game. Examining the extent to which team members influence a player's decision-making process can aid in comprehending the player's motives and distinguishing between active and passive participation in the game.
\item \textit{\textbf{Q3.3}: How do negative player behaviors affect the outcome of a match?} It is imperative to comprehend the impact of negative player conduct on the final result of a match. High-level actors may deliberately concede the game; hence, exploring the association between player behavior and match outcome assumes critical significance.
\end{compactitem}

\textbf{R4: Discern high-level actors within a cohort of players.} After observing the conduct of players during particular matches, experts aim to classify each player. The system should store these classifications/labels and furnish interactive predictions for the remaining unlabeled players. This supportive mechanism mitigates user workload and expedites the labeling process. By repeatedly labeling players, users can gain insights into the attributes of high-level actors and determine the boundaries between ordinary players and high-level actors.
\begin{compactitem}
\item \textit{\textbf{Q4.1}: How to help users label a group of players?} The objective is to facilitate users in labeling a set of players exhibiting comparable behaviors by means of a user-friendly and intuitive interface. The proposed designs are intended to alleviate user burden while also augmenting their analytical capabilities.
\item \textit{\textbf{Q4.2}: How to help users concentrate on a specific group of players during the labeling process?} The system should offer guidance regarding the subsequent target group of players, thereby enabling users to allocate their attention more effectively and advance the labeling task.
\item \textit{\textbf{Q4.3}: How to provide recommendations for users to label high-level actors?} The system should furnish recommendations based on the labeled outcomes of particular players, thereby reducing their workload and affording more time to synthesize the attributes of high-level actors and differentiate them from ordinary players.
\end{compactitem}

\section{{\systemname}}
\par We design {\name} to fulfill the design requirements that pertain to the identification of high-level actors. We first present an overview of the system, followed by an elucidation of the data, events, and metrics employed in the system. Thereafter, individualized introductions will be provided for the various visualization components.

\subsection{System Overview}

\par The system consists of four discrete modules, namely the data processing, interactive model, storage, and visualization modules. The data processing module undertakes the task of processing raw, multivariate data obtained from the MOBA game. It connects this data with specific match and player identifiers and extracts historical matches of specific players while keeping track of which players have played particular heroes. On the other hand, the interactive model module utilizes user-provided labels to recommend labels for remaining unlabeled players. The storage module governs data management and extracts pertinent data when queried by the visualization module. Lastly, the visualization module provides users with the ability to explore both summarized and detailed information within a match and label high-level actors with the assistance of recommendation results. The implementation of the data processing, interactive models, and storage modules is carried out using Python and MongoDB. Meanwhile, the visualization module is developed utilizing Vue.js, Typescript, and D3.

\subsection{Data Abstraction}
\par This study elucidates the various types of raw data that are typically gathered from MOBA games. These data are broadly classified into six categories, namely \textbf{player information}, \textbf{match summary information}, \textbf{multivariate event sequence}, \textbf{multivariate time series data}, \textbf{player movement data}, and \textbf{algorithmic derived data}. To facilitate a deeper understanding of the high-level actors involved in MOBA games, their raw data are subjected to various transformations. We discuss the details of each type of data in {\appendixdata}.

\par The raw data encompasses all essential match-related details for the system. By utilizing the supplied inventory of designated players, it is possible to derive insights concerning the matches of each player, including their historical gameplay, utilizing the game's termination time as a reference. Moreover, it is feasible to generate a summary of information for a particular hero by computing statistics predicated on the hero identifier.

\subsection{Events Abstraction}
\label{sec:events_abstraction}
\par Within the game, players are capable of undertaking a multitude of actions. In order to examine player behavior during the match, it is possible to condense player events on a per-minute basis for the purposes of abstraction and analysis. We detail each kind of event in {\appendixevents}.


\par In this study, we have established a set of events that we refer to as \textbf{priority events}, which are deemed significant occurrences within each minute of gameplay. 
To determine the order of priority for these events, we sought the input of experts in the field. The resulting prioritization sequence is as follows: \textit{turret destruction}, \textit{dragon killing}, \textit{hero killing}, \textit{death}, \textit{assist in killing}, \textit{poke}, \textit{monster killing}, \textit{minion killing}, and \textit{inaction}. It is worth noting that a player may have multiple events occurring within a given minute, but we assign priority to the event with the highest ranking within our established prioritization sequence, designating it as the player's priority event for that specific minute. Furthermore, we have identified a collective event in MOBA games that we refer to as \textbf{team combat}. This event occurs when players from opposing teams are in close proximity and engage in combat. Team combat instances have the potential to greatly affect the tempo of the game and ultimately impact its final outcome.

\subsection{Metrics for Locating Potential High-level Actors}
\par After removing low-level actors through the techniques discussed in {\appendixlowlevel}, it becomes crucial to identify potential high-level actors by employing a collection of metrics. In partnership with experts, we devised and refined the following metrics to rapidly pinpoint potential high-level actors:


\par \textbf{Inactive percentage.} This metric aims to assess a player's positive action and contribution in a MOBA game while considering the possibility of negative behavior and low involvement in certain instances. The metric computation involves determining the percentage of hero damage and economic contribution made by the player within each 20-second interval. The resulting \textit{activeness score} is obtained as the average of these two percentages. If the activeness score is below 0.1, the player is deemed inactive during that interval. The overall inactivity percentage is then calculated as the proportion of time the player remains inactive throughout the game. This metric proves to be beneficial in identifying instances of negative behavior and minimal contribution demonstrated by a player during the game.

\par \textbf{The number of priority events.} To identify potential high-level actors in a MOBA game, we employ a methodology that involves extracting various types of player events (as explained in \autoref{sec:events_abstraction} and {\appendixevents}), followed by computing the frequency of each type of priority events that occur throughout the entire game. We consider inaction events as an important indicator for high-level actors and record the count of such events to assess a player's inactivity during the game. By comparing the distribution of priority events among players, we can obtain a comprehensive summary of the most frequent and significant actions performed by each player during the game. These metrics are instrumental in providing valuable insights into the behavior of individual players and help to identify potential high-level actors.

\subsection{Interactive Models}
\par After a group of players is labeled by users, it becomes necessary to propagate those labels to the remaining players. To accomplish this, we have chosen to employ a robust classification model known as XGBoost~\cite{chen2016xgboost}. The XGBoost model is trained on the current labels and features extracted, as described in {\appendixxgboost}, which allows it to predict a label. These predictions can then aid users in labeling the high-level actors.

\subsection{Visualization}
\par The visualization module comprises four distinct views, namely {\vone}, {\vtwo}, {\vthree}, and {\vfour}. Initially, users are expected to scrutinize {\vone} to obtain an overview of the players of interest (\autoref{fig:teaser}(a)). Afterward, users can choose a specific group of players to investigate further. In {\vtwo}, a summary of the behavioral patterns of the selected group of players is presented (\autoref{fig:teaser}(b)). Users may proceed to examine {\vthree} to select a particular player of interest (\autoref{fig:teaser}(c)). Subsequently, {\vtwo} may be utilized to display the summary information of the behavior patterns in the historical matches of the selected player or other players' matches involving the same hero (\autoref{fig:teaser}(b)). {\vfour} displays key events and trajectories of all ten players in a single match, facilitating users' comprehension of the game's dynamics (\autoref{fig:teaser}(d)). These investigations help users in making informed judgments regarding whether the selected player is a high-level actor. Users may return to {\vone} to label the selected players, and the system can generate recommendations for labeling the remaining players (\autoref{fig:teaser}(a)).

\subsubsection{\vone}
\par In order to facilitate the rapid identification of potential high-level actors among a group of players, {\vone} has been devised to enable users to efficiently comprehend the players' statuses across multiple matches (\autoref{fig:teaser}(a)) (\textbf{R1}, \textbf{R4}). This view is composed of three distinct panels, namely the \textit{Statistics panel}, \textit{Filters panel}, and \textit{Projection panel}. The Statistics panel presents pertinent information regarding the number of players that have been focused and labeled (\autoref{fig:teaser}(a1)). The Filters panel enables users to apply various filters to the focused players, such as by their labeling status or percentage of inactivity (\autoref{fig:teaser}(a2)). The Projection panel employs glyphs to depict one player in one match, furnishing users with a synopsis of the focused players in the current session (\autoref{fig:teaser}(a3)).

  \begin{figure}[h]
  \centering
  \vspace{-3mm}
  \includegraphics[width=\linewidth]{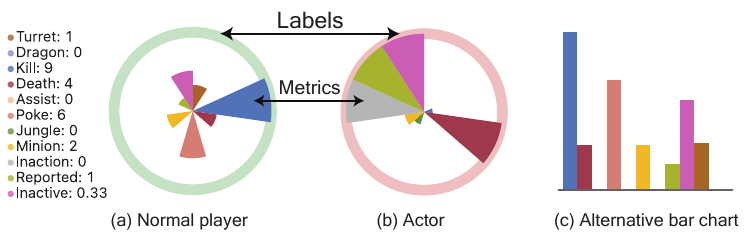}
    \vspace{-6mm}
  \caption{The glyph in {\vone} is used to represent the metrics of one player in one match. (a) When the players are labeled as normal, the border of the outer circle is colored green. (b) When the players are labeled as actors, the border of the outer circle is colored red. (c) An alternative design of the glyph is presented.}
  \label{fig:glyph_design}
    \vspace{-3mm}
\end{figure}  

\par \textbf{Visual Design.} The glyph representation incorporates eleven distinct metrics, encompassing the number of nine priority events, the inactive percentage, and the number of reports attributed to the player (\autoref{fig:glyph_design}). The outer circle's border color of the glyph denotes the player's current label, with green used for normal players (\autoref{fig:glyph_design}(a)) and red for actors (\autoref{fig:glyph_design}(b)), while the outer circle's border thickness increases if the player has already been labeled. If a player has not been labeled, the outer circle's border color encodes the prediction labels. \mzhihua{Since the UMAP projection is fast and reliable, the UMAP projection with collision avoidance technique is employed to position the glyphs with comparable metrics together~\cite{mcinnes2018umap}.} The distance between the glyphs is computed by summing the squares of the discrepancies among the normalized metrics.

\par \textbf{Interaction.} Users have the ability to lasso select a cluster of players for closer examination. Within the Projection panel, users can label players and employ interactive models to generate label suggestions for any remaining unlabeled players. Subsequently, users can download the labeled outcomes for the players to conduct further analysis or train a classifier aimed at distinguishing between these players.

\par \textbf{Design Alternative.} Numerous alternative designs for the glyph representation were evaluated, including the utilization of a bar chart to depict the variables (\autoref{fig:glyph_design}(c)). However, this design was ultimately disregarded since it could pose a challenge for users to identify and compare the bars with other glyphs when the glyph's size is limited within the Projection panel. Consequently, the current glyph design was selected.

\subsubsection{\vtwo}
\par
\begin{figure}[h]
  \centering
    \vspace{-3mm}
  \includegraphics[width=\linewidth]{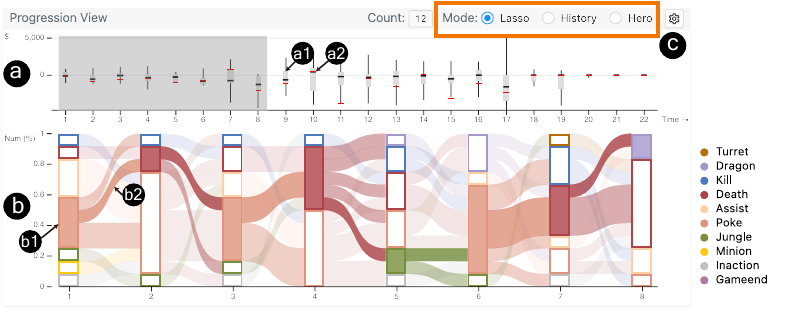}
    \vspace{-6mm}
  \caption{{\vtwo} offers a concise summary of the behaviors exhibited by a group of players across multiple matches. Specifically, (a) a box plot is utilized to illustrate the typical range of economic differences within the group, while (b) the vertical bar represents the distribution of priority events, and the flow indicates the transition between these events over consecutive timestamps.}
  \label{fig:progression_view}
    \vspace{-3mm}
\end{figure}

 {\vtwo} is designed to offer a comprehensive overview of the activities, advancement patterns, and connection between activities and economic disparities of a group of players across multiple matches, with the goal of obtaining insights into their behavior (\autoref{fig:progression_view}) (\textbf{R2}, \textbf{R4}). It enables users to swiftly comprehend the economic disparity and priority events distribution in a group of players during a specific time period. The concept of priority event is introduced in \autoref{sec:events_abstraction}.

\begin{figure*}[h]
  \centering
    \vspace{-3mm}
  \includegraphics[width=\linewidth]{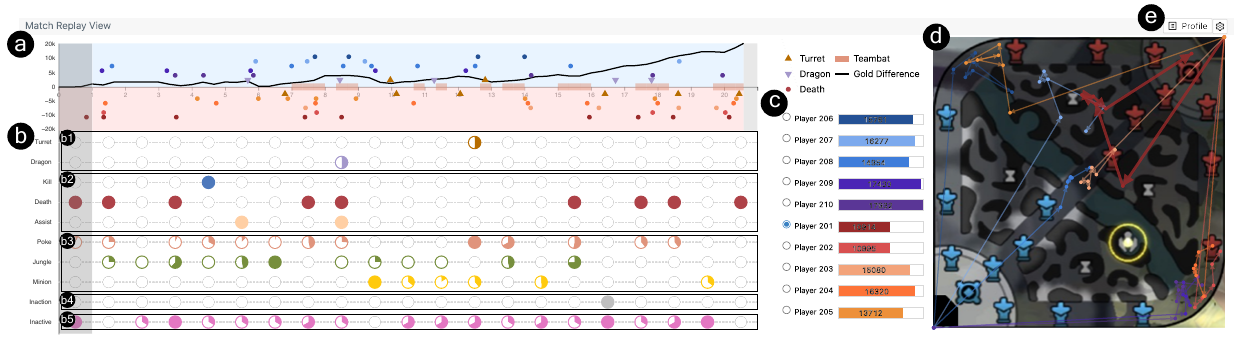}
    \vspace{-6mm}
  \caption{{\vfour} offers a concise representation of the game's battle progress. Specifically, it comprises four distinct chart types: (a) the match summary chart, which depicts the gold difference and significant events that occurred during a single match between two teams; (b) the player events chart, which displays the events that occurred in each minute of the game for the focal player; (c) the player summary chart, which compares the economies of multiple players over the course of the match; and (d) the map chart, which shows the trajectories of all ten players during a specific time range of the game.}
  \label{fig:match_replay_view}
    \vspace{-6mm}
\end{figure*}

\par \textbf{Visual Design.} The upper section of the {\vtwo} provides a minute-by-minute display of the economic difference between a group of players and a selected player in the {\vthree} (\autoref{fig:progression_view}(a)). To depict the normal range of economic differences among concerned players, a box plot is used (\autoref{fig:progression_view}(a1)). To aid users in comprehending the economic difference of the currently selected player, a red line is displayed at each minute (\autoref{fig:progression_view}(a2)). Users have the option to select the region where the economic difference is negative, indicating an unfavorable economic situation, which may suggest negative behaviors during that time period. The lower section of the {\vtwo} illustrates the distribution of priority events and their transition patterns (\autoref{fig:progression_view}(b)). The height of each vertical bar indicates the percentage of players who performed each priority event in the group (\autoref{fig:progression_view}(b1)). The priority event type is denoted by the color of the bar's stroke, and the width of the flow between two bars represents the percentage of players who consecutively performed the two priority events in two nearby minutes (\autoref{fig:progression_view}(b2)). If the group's percentage of low-ranked priority events, such as monster killing, minion killing, and inaction, is low, it is more likely that the player experienced abnormal events during that minute. If a player is selected in the {\vthree}, the bar and flow of that player's specific priority events are highlighted.

\par \textbf{Interaction.} Users have the ability to filter players based on specific criteria by selecting a bar or flow with a particular priority event during certain timestamps. 
\mzhihua{The user can also modify the cohort of players under investigation, including those selected in the Projection panel (lasso mode), a player's past matches (history mode), and matches of other players who have played the same hero (hero mode) (\autoref{fig:progression_view}(c)).}
Analysis of actions can be carried out in three dimensions, which are their commonality among the selected group of players, their prevalence in a player's previous matches, and their occurrence in matches of other players who have played the same hero. 

\subsubsection{\vthree}

\par The {\vthree} is designed to help users focus on a particular player and match by summarizing the player's performance in the match (\autoref{fig:teaser}(c))~(\textbf{R3}, \textbf{R4}). The table contains multiple rows, each presenting different information about one player, including KDA (kills plus assists divided by deaths plus one) and inactive percentage. These data can be helpful for users to focus on a specific player. A histogram of the metrics is shown at the top of the table to give users an idea of the metric distribution among the chosen players (\autoref{fig:teaser}(c1)). If a group of players is selected in {\vone} by lassoing, {\vthree} will only display information about the chosen players. To select a particular player in a match, users can click the radio button next to the row. {\vone} and {\vtwo} will highlight the selected player, while {\vfour} will display the essential events of the game for that player. Users can give a label for that player at the top of this view (\autoref{fig:teaser}(c2)).

\subsubsection{\vfour}
\par To facilitate the user's understanding of essential events and anomalous time periods in a particular match, {\vfour} provides a comprehensive overview of the entire match (\autoref{fig:match_replay_view}) (\textbf{R3}, \textbf{R4}). The view consists of four charts: the \textit{match summary chart}, \textit{player events chart}, \textit{player summary chart}, and \textit{map chart}. These charts are combined to reconstruct the battle progress and assist users in comprehending the selected player's significant events during the match, enabling them to identify specific abnormal behaviors at particular timestamps.


\par \textbf{Visual Design.} The match summary chart, depicted in~\autoref{fig:match_replay_view}(a), offers a concise representation of crucial events and the economic difference in a match over time. The timing of the game is plotted on the x-axis, while the y-axis represents the accumulated economic difference. The chart employs various symbols to represent important events, such as the players' death shown as a circle, the destruction of the enemy team's turret represented by an up-pointing triangle, and the team's killing of the dragon indicated by a down-pointing triangle. The rectangle on the x-axis denotes the duration of the team combat. This chart provides users with a quick understanding of the current battle progress and the occurrence of key events.


\par The player events chart, as presented in \autoref{fig:match_replay_view}(b), provides a concise summary of the selected player's events throughout the game. Each row, except for the last row, corresponds to the events that took place during each minute of the game. If an event occurred during the given time interval, the stroke color of the circle changes to reflect the event's color. For turret destruction and dragon killing events, if the player contributed to the damage but did not deliver the killing blow, a semicircle fills the entire circle (\autoref{fig:match_replay_view}(b1)). The percentage of poke, monster killing, and minion killing events represents the damage inflicted on the opponent's hero and the economy received from these events. These metrics are normalized by the maximum value across the game time for that player, as shown in \autoref{fig:match_replay_view}(b3). In the case of kill, death, assist, and inaction events, if they occurred during the corresponding time interval, the circle is fully filled (\autoref{fig:match_replay_view}(b2, b4)). The last row depicts the inactive status of the player throughout the game (\autoref{fig:match_replay_view}(b5)). The percentage of the inactive circle represents the amount of inactive time in that minute. Such summaries assist users in comprehending the detailed events of each minute for that player.

\par The player summary chart exhibits the economic status of each player in the game (\autoref{fig:match_replay_view}(c)). The width of each bar indicates the economic value of the corresponding player, and a distinct color is assigned to each player for identification. This chart enables users to compare the economic performance of a particular player with that of other players and identify any unusual economic behaviors.

\par The map chart in the {\vfour} displays the movements of the ten players within a brushed time range (\autoref{fig:match_replay_view}(d)). The color of each trajectory represents the respective player, while the selected player's trajectory is represented with a thicker line. Analyzing the movement patterns of the players enables users to identify the actions taken during a particular time frame and assess their strategic significance.


\par \textbf{Interaction.}  Users have the option to select a specific time period in the match summary chart or player events chart to investigate the trajectories of players during that time period. Additionally, if users require further information regarding the profile of players, they may click the Profile button located in the view header (\autoref{fig:match_replay_view}(e)). Furthermore, users can select a particular player in the match by clicking the radio button present in the player summary chart to conduct a more detailed analysis.
\begin{figure*}[h]
\centering
  \vspace{-3mm}
\includegraphics[width=\linewidth]{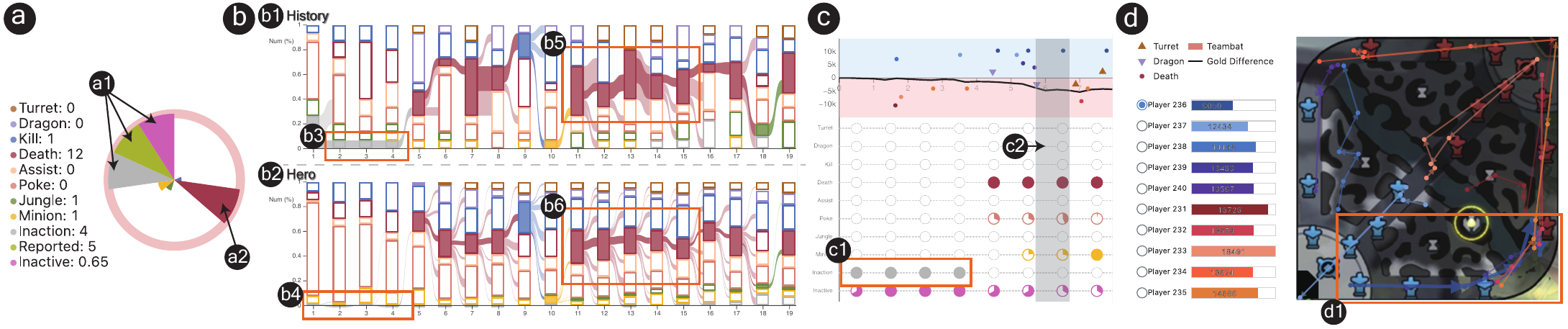}
  \vspace{-6mm}
\caption{This case serves as evidence that E2 is capable of identifying instances where a player exhibits AFK (Away From Keyboard) behavior and intentionally incurs death by the turret. Specifically, this is supported by (a) the glyph's depiction of players displaying multiple anomalous metrics, (b) the history and hero mode of {\vtwo} revealing numerous low percentage priority events among players, and (c, d) the identification of a particular player rushing towards enemy turrets and ultimately meeting his demise as indicated in the map chart of {\vfour}.}
\label{fig:case_1_1}
  \vspace{-6mm}
\end{figure*}

\section{Evaluation}
\par Two cases are presented to demonstrate how the experts explored and identified high-level actors among a group of players. Case one is described in \autoref{sec:case_one}, and case two is presented in {\appendixcasetwo}. Subsequently, a user study was carried out with 12 experts to evaluate the effectiveness and usability of the system. Detailed descriptions and analysis results of the user study are presented in \autoref{sec:user_study}.

\subsection{AFK Behavior, Intentional Turret Kill, and Non-Participation in Dragon Killing}
\label{sec:case_one} 
\par E2 specializes in identifying high-level actors in MOBA games through the development of algorithms and possesses a deep interest in analyzing the behavior of such actors within the context of MOBA games.

\par \textbf{AFK and intentionally killed by the turret.} Initially, E2 observed that the number of players under consideration was 30, as displayed in the Statistics panel of {\vone} (\autoref{fig:teaser}(a1)). Subsequently, E2 refined the selection by applying filters in the Filters panel of {\vone} to focus exclusively on players with a number of reports ranging from 3 to 5 and an inactive percentage ranging from 0.50 to 0.65 (\autoref{fig:teaser}(a2)). Next, E2 identified one player from the Projection panel who had the highest number of reports, inaction, inactive percentage (\autoref{fig:case_1_1}(a1)), and deaths (\autoref{fig:case_1_1}(a2)) among the selected group of players (\autoref{fig:case_1_1}(a)). The metrics of this player raised suspicion, suggesting that the player exhibited negative gameplay behavior most of the time (\textbf{R1}).

\par Further analysis of the player's history (\autoref{fig:case_1_1}(b1)) and hero mode (\autoref{fig:case_1_1}(b2)) in the {\vtwo} revealed that the player had inaction events for the first four minutes (\autoref{fig:case_1_1}(b3, b4)) and a high number of deaths for the remaining time (\autoref{fig:case_1_1}(b5, b6)). Notably, the percentage of inaction events in the beginning was low, indicating that the player's behavior was atypical and unlikely to be exhibited by other players playing the same hero or in his previous games. Moreover, the low probability of transitioning between two death events suggested that the player was more likely to be negatively engaged in the game (\textbf{R2}).

\par To gain further insight, E2 examined the trajectory of the player at the beginning of the game in the {\vfour} and observed that the player stayed in the highland (\autoref{fig:case_1_1}(c1)) before intentionally rushing towards the turret and getting killed (\autoref{fig:case_1_1}(c2, d1)) (\textbf{R3}). This raised the question of why low-level detection algorithms (\appendixlowlevel) failed to detect such an apparent case of negative gameplay behavior. To address this, E2 investigated the player's profile information by clicking the Profile button in the {\vfour} and found that the idle time was only 111 seconds, below the threshold of 120 seconds for the AFK actor detection algorithm. Based on this analysis, E2 concluded that combining the identification of prolonged periods of inactivity with a high number of death events would enable the identification of such negative gameplay behavior. Finally, E2 labeled the player as an actor in the {\vone} (\textbf{R4}).

\par \textbf{Non-participation in dragon killing.} E2 utilized lasso selection to identify the remaining four players in the Projection Panel (\autoref{fig:teaser}(a4)) (\textbf{R1}). Subsequently, E2 examined the {\vtwo} in lasso mode and found that the economic disadvantage is relatively low during the $12^{th}$, $13^{th}$, and $15^{th}$ minutes (\autoref{fig:teaser}(b1)), which implies that the disadvantages accumulate during those time periods. After brushing those time periods, E2 observed that one player had multiple low-ranked priority events in consecutive timestamps (\autoref{fig:teaser}(b2)) (\textbf{R2}). E2 then proceeded to select the flow between minion-killing events during the $14^{th}$ and $15^{th}$ minutes and identified the player as 243 in match 25. E2 found that the player's KDA in {\vthree} was zero (\autoref{fig:teaser}(c3)), and abnormal events persisted during the $12^{th}$ to $15^{th}$ minute in both history and hero modes of the {\vtwo}. 
Upon investigating the {\vfour}, it was discovered that during the $14^{th}$ to $15^{th}$ minute, the player solely focused on his own lane (\autoref{fig:teaser}(d4)), resulting in the failure of the team's efforts to defeat the dragon (\autoref{fig:teaser}(d5)), with all of the player's teammates dying during the combat (\autoref{fig:teaser}(d3)) (\textbf{R3}). This behavior suggests a high likelihood of the player being an actor. As a result, E2 labeled the player as an actor in the Projection Panel (\textbf{R4}).

\subsection{User Study}
\label{sec:user_study}

\subsubsection{Participants}
\par In this study, we utilized email to recruit 12 experts (age: 21-32, $\mu=25.42$, $\sigma=3.65$), including two experts in actor identification (E1-E2) and ten with extensive experience in playing MOBA games (E4-E13). The actor identification experts (E1-E2) were selected based on their involvement in conducting actor identification in a leading game company, as described in \autoref{sec:abstractions}. Given the scarcity of experts in this domain, we sought additional experts with expertise in MOBA game scenarios and gameplay to supplement our study. Therefore, we included ten additional experts (E4-E13) with significant knowledge and experience in playing MOBA games in our recruitment efforts.


\subsubsection{Experiment Design}
\par In order to assess the system's ability to identify and label high-level actors and evaluate its usability, we carried out a user study involving 12 experts.

\par \textbf{Baseline.} The original protocol for inspecting games at a game company entails the use of video checking to label actors, a task that can be time-consuming even when performed at a rapid pace. Based on the interview records with experts, labeling one player in a single match can take eight minutes on average. As the participants re-watch the video is time-consuming, it is not affordable to run such experiments. Therefore, we compare the duration reported by the experts in this study to evaluate whether the usage of {\name} accelerates the inspection process. We consulted the experts on whether the accuracy of the usage of {\name} meets their expectations.

\par \textbf{Data and tasks.} A labeled MOBA game dataset containing information on whether a player is an actor has been collected from a game company and serves as the ground truth for this study. The labels were generated by a game inspector through video checking. In order to filter out low-level actors, the low-level actor detection algorithms ({\appendixlowlevel}) were initially employed. Subsequently, two datasets were constructed by randomly sampling from the full dataset. The first dataset, consisting of 10 normal players and 10 actors, was created for the purpose of familiarizing participants with the system. The second test dataset included 30 normal players and 30 actors. To ensure that the experiments were conducted within a reasonable timeframe, we randomly sampled 15 normal players and 15 actors from the test dataset to conduct each experiment. Participants were asked to label these players based on whether they were normal or actors.

\par \textbf{Procedures.} Initially, a comprehensive introduction was provided to the system interface and the workflow for utilizing the system, which was presented for a duration of 30 minutes. To facilitate comprehension of the system, cases involving the detection of feeders and actors who were not involved in the main events were demonstrated, emphasizing the critical visual cues required for exploring the system. Subsequently, participants were requested to identify and label one normal player and one high-level actor from the first demo dataset, utilizing the system, which also took 30 minutes. Following this, participants were requested to label the test datasets using the system. Finally, participants were provided with a post-questionnaire to provide qualitative and quantitative feedback on the system's effectiveness and usability, as well as suggestions for system improvement. Participants were also instructed to upload their labeling outcomes for further analysis. On average, the experiments were completed within two hours.

\subsubsection{Results and Analysis}

\par \noindent\textbf{Accuracy and duration.} In this study, we have evaluated the accuracy and duration of participants using the system to label players as either normal players or actors. 
The accuracy for using the system to label players is $74.17 \pm 7.12$, and the duration is $25.92 \pm 10.48$ minutes.
In terms of duration, we found that our system can accelerate this process compared to traditional workflow, which requires 52 seconds on average for labeling one player in one match (10.8\% of the time needed to label a single match in the traditional workflow). We consulted the experts from the leading game company in terms of performance. They commented that the current demonstrated performance meets their expectations, given that high-level actor identification is quite difficult.

\par \noindent\textbf{Effectiveness.} The majority of experts (9/12) expressed their satisfaction with the system's capability to facilitate the identification and accurate labeling of high-level actors. Specifically, experts E1-E2, E4-E6, and E8-E9 acknowledged the usefulness of {\vfour} in actor identification, with E2, E4-E5, and E8-E9 further emphasizing the value of player trajectories displayed on the map chart in helping them differentiate between players who lack proficiency in the game and those intentionally undermining game fairness. E5 and E8 also stated that ``\textit{checking trajectories in the map}'' enables them to determine whether a player is actively participating in team combat. In contrast, E7 appreciated the grouping of similar players in {\vone}, which allows for the simultaneous analysis of multiple players, along with the recommendation system that helps in reducing the labeling workload. 
E13 further noted that the summary information on the behaviors of players who play the same hero as the focal player enables him to determine whether the player intentionally failed or just followed the mainstream actions. 


\par \noindent\textbf{Visual design.} \mzhihua{Experts appreciated the usefulness of the current design.} E1, E2, E6, and E12 appreciated the design of {\vfour} for its ability to provide a quick overview of events across multiple matches. E5, E8, and E13 found {\vtwo} and {\vthree} particularly useful for filtering players based on specific behaviors and timestamps. However, some limitations were identified by the experts, with E5 commenting that ``\textit{if the color for players in {\vfour} can represent the assigned location or hero type, it will be better to check whether the player is on the right lane.}'' E1 commented that \textit{\dq{if the {\vfour} can tightly connect the key events and players' actions, it would be better to identify high-level actors.}}

\noindent\textbf{Usability.} Most of the participants (9/12) appreciated that the system would be useful for detecting high-level actors and would like to recommend others who are working on actor identification to explore the system. E6 and E8 expressed that through filtering by the number of reports, users can divide the players into multiple groups and further inspect them individually. However, E1, E7, and E10 also commented that \textit{\dq{the interaction between views is quite complex, and the learning curve will be high for novice users.}}

\par \noindent\textbf{Suggestions.} The experts provided several suggestions for improving the design and future directions of the system. 
E5 and E11 suggested that it would be better to have a global overview of trajectories in the {\vfour} rather than only focusing on a small piece of the time period in each brush interaction. E1, E5, and E6 suggested presenting more details to assist users in understanding the reasons for each death event, such as displaying the trajectories leading up to the death events and the death location on the map. E7 suggested that, in terms of the recommendation system, it would be beneficial to provide more information about the reasons for its prediction and the player's similarity to the labeled players.


\section{Discussion and Limitations}

\noindent\textbf{Lessons learned.} Through this design study, we have gained valuable insights that can potentially be generalized to other similar design study papers. Firstly, our findings suggest that when investigating the intention behind inspected behaviors, it is necessary to compare the presented behaviors with past observed behaviors of actors that are stored in the inspector's memory. Therefore, presenting and depicting players' behaviors in detail could serve as a useful aid in refreshing the inspector's working memory and assisting their judgment. Secondly, our initial plan was to cluster a large number of players and present a summary of players in each cluster to assist in labeling. However, after consulting with experts, it became apparent that the behavior of players in a MOBA game is quite complex and heavily influenced by other players. Hence, in scenarios where the focused players have complex event sequences in the match, narrowing down to a single match and contextualizing it with other matches can be a more effective approach to help users understand the events that occurred.


\noindent\textbf{Generalizabiliy and scalability.} Our system is dependent on the event abstraction of the particular MOBA game. However, it can be potentially extended to other MOBA games as long as the event abstraction is similar. \mzhihua{Beyond MOBA games, it can be beneficial to other collaboration and competition fields like this scenario. For example, there may be match-fixing in football games. We can generalize our techniques and summarize their events to find the abnormal events in the progress of the game. }
Nonetheless, scalability is a critical concern for {\vone}, as it is currently limited to a maximum of 300 players. When handling more than $300$ players in multiple matches, it will be necessary to filter the players and divide them into smaller groups to facilitate inspection and labeling.

\noindent\textbf{Limitations.} Limitations exist that must be taken into consideration. First, we prioritize a list of events, leading to the hiding of other events when multiple events occur simultaneously. This approach may result in some events being missed by the user during the inspection of the {\vone} and {\vtwo}. Second, the interactions between the different views can be complex and confusing for the users. Based on user feedback, it was found that users preferred labeling one player in one match at a time when confirming behaviors in {\vfour}. To address this issue, we added a labeling function in {\vthree} to enable labeling of one player in one match. \mzhihua{Third,  the interactive models can indeed mitigate the workload of users for labeling players. However, the training of interactive models requires many training instances while the number of given labels is typically small. Thus, the accuracy of interactive models may not be good enough to assist users in providing reliable labels.  Fourth, since the system is targeting domain experts, the design of visualization should be simple and intuitive. Common visualization is chosen to mitigate their learning curve and workload, although it is still very hard for them to understand them, and they think that the learning curve is high. }

\section{Conclusion and Future Work}

\par MOBA games have attracted players worldwide; however, certain individuals may engage in actions that compromise the fairness of the game. To facilitate the identification and labeling of high-level actors, we have developed a visual analytics system named {\name}. This system is comprised of four distinct views, namely {\vone}, {\vtwo}, {\vthree}, and {\vfour}. The {\vone} is responsible for grouping similar players and filtering players to provide users with an overview of player status across matches. The {\vtwo} summarizes match behaviors across three dimensions, including lasso-selected players, the historical matches of the focused player, and the matches of other players playing the same hero. The {\vthree} provides a summary of player information and enables further investigation of a specific player. Finally, the {\vfour} allows users to inspect detailed player behaviors and trajectories in the map, enabling the determination of concrete abnormal events and timestamps. We conducted case studies with two experts and user studies with 12 experts to demonstrate the effectiveness and usability of the system.

\par In light of the results of the user study and limitations identified during system usage, we aim to implement additional measures to address the issues identified in the future. Our proposed solutions comprise the following actions. First, we intend to enhance the color encoding options available to users, enabling them to select the precise meaning of color encoding for players. This could involve representing assigned locations or hero types using specific color codes, among other possibilities. Second, we aim to improve the filtering algorithm, given that matches requiring inspection are often large, to increase recall and reduce the workload of human users. Lastly, we plan to incorporate more detailed game-specific information in {\vfour}, such as death location in death events, a replay of trajectories leading up to death, and a summary of reasons for a player's death.


\bibliographystyle{abbrv-doi-hyperref}

\bibliography{main, modpaperpipe}

\begin{thebibliography}{10}

\bibitem{Adinolf2018-wf}
S.~Adinolf and S.~Turkay.
\newblock Toxic behaviors in esports games: Player perceptions and coping
  strategies.
\newblock In {\em Proceedings of the Annual Symposium on {Computer-Human}
  Interaction in Play Companion Extended Abstracts}, pp. 365--372. {ACM}, 2018.
  \href{https://doi.org/10.1145/3270316.3271545}
{doi: {{%
10\hspace{.1pt}\discretionary{.}{%
}{.}\hspace{.4pt}1145\discretionary{/}{%
}{/}3270316\hspace{.1pt}\discretionary{.}{%
}{.}\hspace{.4pt}3271545}}}


\bibitem{Afonso2019-of}
A.~P. Afonso, M.~B. Carmo, T.~Gonçalves, and P.~Vieira.
\newblock {VisuaLeague}: Player performance analysis using spatial-temporal
  data.
\newblock {\em Multimedia Tools and Applications}, 78(23):33069--33090, 2019.
  \href{https://doi.org/10.1007/s11042-019-07952-z}
{doi: {{%
10\hspace{.1pt}\discretionary{.}{%
}{.}\hspace{.4pt}1007\discretionary{/}{%
}{/}s11042\discretionary{%
}{-}{-}019\discretionary{%
}{-}{-}07952\discretionary{%
}{-}{-}z}}}


\bibitem{Ahmad2019-uq}
S.~Ahmad, A.~Bryant, E.~Kleinman, Z.~Teng, T.-H.~D. Nguyen, and M.~Seif
  El-Nasr.
\newblock Modeling individual and team behavior through spatio-temporal
  analysis.
\newblock In {\em Proceedings of the Annual Symposium on {Computer-Human}
  Interaction in Play}, pp. 601--612. {ACM}, 2019.
  \href{https://doi.org/10.1145/3311350.3347188}
{doi: {{%
10\hspace{.1pt}\discretionary{.}{%
}{.}\hspace{.4pt}1145\discretionary{/}{%
}{/}3311350\hspace{.1pt}\discretionary{.}{%
}{.}\hspace{.4pt}3347188}}}


\bibitem{Ashton2011-uc}
M.~Ashton and C.~Verbrugge.
\newblock Measuring cooperative gameplay pacing in world of warcraft.
\newblock In {\em Proceedings of the 6th International Conference on
  Foundations of Digital Games}, pp. 77--83. {ACM}, 2011.
  \href{https://doi.org/10.1145/2159365.2159376}
{doi: {{%
10\hspace{.1pt}\discretionary{.}{%
}{.}\hspace{.4pt}1145\discretionary{/}{%
}{/}2159365\hspace{.1pt}\discretionary{.}{%
}{.}\hspace{.4pt}2159376}}}


\bibitem{Bay2003-hu}
S.~D. Bay and M.~Schwabacher.
\newblock Mining distance-based outliers in near linear time with randomization
  and a simple pruning rule.
\newblock In {\em Proceedings of the Ninth {ACM} {SIGKDD} International
  Conference on Knowledge Discovery and Data Mining}, pp. 29--38, 2003.
  \href{https://doi.org/10.1145/956750.956758}
{doi: {{%
10\hspace{.1pt}\discretionary{.}{%
}{.}\hspace{.4pt}1145\discretionary{/}{%
}{/}956750\hspace{.1pt}\discretionary{.}{%
}{.}\hspace{.4pt}956758}}}


\bibitem{Breunig2000-fl}
M.~M. Breunig, H.-P. Kriegel, R.~T. Ng, and J.~Sander.
\newblock {LOF}: Identifying density-based local outliers.
\newblock In {\em Proceedings of the {ACM} {SIGMOD} International Conference on
  Management of Data}, pp. 93--104, 2000.
  \href{https://doi.org/10.1145/342009.335388}
{doi: {{%
10\hspace{.1pt}\discretionary{.}{%
}{.}\hspace{.4pt}1145\discretionary{/}{%
}{/}342009\hspace{.1pt}\discretionary{.}{%
}{.}\hspace{.4pt}335388}}}


\bibitem{bryan2014isse}
N.~J. Bryan, G.~J. Mysore, and G.~Wang.
\newblock Isse: An interactive source separation editor.
\newblock In {\em Proceedings of the SIGCHI Conference on Human Factors in
  Computing Systems}, pp. 257--266, 2014.
  \href{https://doi.org/10.1145/2556288.2557253}
{doi: {{%
10\hspace{.1pt}\discretionary{.}{%
}{.}\hspace{.4pt}1145\discretionary{/}{%
}{/}2556288\hspace{.1pt}\discretionary{.}{%
}{.}\hspace{.4pt}2557253}}}


\bibitem{Canossa2021-vu}
A.~Canossa, D.~Salimov, A.~Azadvar, C.~Harteveld, and G.~Yannakakis.
\newblock For honor, for toxicity: Detecting toxic behavior through gameplay.
\newblock {\em Proceedings of the ACM on Human-Computer Interaction}, 5:1--29,
  2021. \href{https://doi.org/10.1145/3474680}
{doi: {{%
10\hspace{.1pt}\discretionary{.}{%
}{.}\hspace{.4pt}1145\discretionary{/}{%
}{/}3474680}}}


\bibitem{Cao2017-ki}
N.~Cao, C.~Lin, Q.~Zhu, Y.~Lin, X.~Teng, and X.~Wen.
\newblock Voila: Visual anomaly detection and monitoring with streaming
  spatiotemporal data.
\newblock {\em IEEE Transactions on Visualization and Computer Graphics},
  24(1):23--33, 2018. \href{https://doi.org/10.1109/TVCG.2017.2744419}
{doi: {{%
10\hspace{.1pt}\discretionary{.}{%
}{.}\hspace{.4pt}1109\discretionary{/}{%
}{/}TVCG\hspace{.1pt}\discretionary{.}{%
}{.}\hspace{.4pt}2017\hspace{.1pt}\discretionary{.}{%
}{.}\hspace{.4pt}2744419}}}


\bibitem{Cao2016-iq}
N.~Cao, C.~Shi, S.~Lin, J.~Lu, Y.-R. Lin, and C.-Y. Lin.
\newblock {TargetVue}: Visual analysis of anomalous user behaviors in online
  communication systems.
\newblock {\em IEEE Transactions on Visualization and Computer Graphics},
  22(1):280--289, 2016. \href{https://doi.org/10.1109/TVCG.2015.2467196}
{doi: {{%
10\hspace{.1pt}\discretionary{.}{%
}{.}\hspace{.4pt}1109\discretionary{/}{%
}{/}TVCG\hspace{.1pt}\discretionary{.}{%
}{.}\hspace{.4pt}2015\hspace{.1pt}\discretionary{.}{%
}{.}\hspace{.4pt}2467196}}}


\bibitem{Chae2012-wa}
J.~Chae, D.~Thom, H.~Bosch, Y.~Jang, R.~Maciejewski, D.~S. Ebert, and T.~Ertl.
\newblock Spatiotemporal social media analytics for abnormal event detection
  and examination using seasonal-trend decomposition.
\newblock In {\em {IEEE} Conference on Visual Analytics Science and
  Technology}, pp. 143--152, 2012.
  \href{https://doi.org/10.1109/VAST.2012.6400557}
{doi: {{%
10\hspace{.1pt}\discretionary{.}{%
}{.}\hspace{.4pt}1109\discretionary{/}{%
}{/}VAST\hspace{.1pt}\discretionary{.}{%
}{.}\hspace{.4pt}2012\hspace{.1pt}\discretionary{.}{%
}{.}\hspace{.4pt}6400557}}}


\bibitem{Chandola2009-wv}
V.~Chandola, A.~Banerjee, and V.~Kumar.
\newblock Anomaly detection: A survey.
\newblock {\em ACM Computing Surveys}, 41(3):1--58, 2009.
  \href{https://doi.org/10.1145/1541880.1541882}
{doi: {{%
10\hspace{.1pt}\discretionary{.}{%
}{.}\hspace{.4pt}1145\discretionary{/}{%
}{/}1541880\hspace{.1pt}\discretionary{.}{%
}{.}\hspace{.4pt}1541882}}}


\bibitem{Charleer2018-kv}
S.~Charleer, K.~Gerling, F.~Gutiérrez, H.~Cauwenbergh, B.~Luycx, and
  K.~Verbert.
\newblock {Real-Time} dashboards to support esports spectating.
\newblock In {\em Proceedings of the Annual Symposium on {Computer-Human}
  Interaction in Play}, pp. 59--71. {ACM}, 2018.
  \href{https://doi.org/10.1145/3242671.3242680}
{doi: {{%
10\hspace{.1pt}\discretionary{.}{%
}{.}\hspace{.4pt}1145\discretionary{/}{%
}{/}3242671\hspace{.1pt}\discretionary{.}{%
}{.}\hspace{.4pt}3242680}}}


\bibitem{chen2016xgboost}
T.~Chen and C.~Guestrin.
\newblock Xgboost: A scalable tree boosting system.
\newblock In {\em Proceedings of the 22nd {ACM} {SIGKDD} International
  Conference on Knowledge Discovery and Data Mining}, pp. 785--794, 2016.
  \href{https://doi.org/10.1145/2939672.2939785}
{doi: {{%
10\hspace{.1pt}\discretionary{.}{%
}{.}\hspace{.4pt}1145\discretionary{/}{%
}{/}2939672\hspace{.1pt}\discretionary{.}{%
}{.}\hspace{.4pt}2939785}}}


\bibitem{Cheng2019-sm}
Z.~Cheng, Y.~Yang, C.~Tan, D.~Cheng, A.~Cheng, and Y.~Zhuang.
\newblock What makes a good team? {A} large-scale study on the effect of team
  composition in honor of kings.
\newblock In {\em Proceedings of the World Wide Web Conference}, pp.
  2666--2672. {ACM}, 2019. \href{https://doi.org/10.1145/3308558.3313530}
{doi: {{%
10\hspace{.1pt}\discretionary{.}{%
}{.}\hspace{.4pt}1145\discretionary{/}{%
}{/}3308558\hspace{.1pt}\discretionary{.}{%
}{.}\hspace{.4pt}3313530}}}


\bibitem{cui2007easyalbum}
J.~Cui, F.~Wen, R.~Xiao, Y.~Tian, and X.~Tang.
\newblock Easyalbum: An interactive photo annotation system based on face
  clustering and re-ranking.
\newblock In {\em Proceedings of the SIGCHI Conference on Human Factors in
  Computing Systems}, pp. 367--376. {ACM}, 2007.
  \href{https://doi.org/10.1145/1240624.1240684}
{doi: {{%
10\hspace{.1pt}\discretionary{.}{%
}{.}\hspace{.4pt}1145\discretionary{/}{%
}{/}1240624\hspace{.1pt}\discretionary{.}{%
}{.}\hspace{.4pt}1240684}}}


\bibitem{deng2014scalable}
J.~Deng, O.~Russakovsky, J.~Krause, M.~S. Bernstein, A.~Berg, and L.~Fei-Fei.
\newblock Scalable multi-label annotation.
\newblock In {\em Proceedings of the SIGCHI Conference on Human Factors in
  Computing Systems}, pp. 3099--3102. {ACM}, 2014.
  \href{https://doi.org/10.1145/2556288.2557011}
{doi: {{%
10\hspace{.1pt}\discretionary{.}{%
}{.}\hspace{.4pt}1145\discretionary{/}{%
}{/}2556288\hspace{.1pt}\discretionary{.}{%
}{.}\hspace{.4pt}2557011}}}


\bibitem{Drachen2009-rq}
A.~Drachen and A.~Canossa.
\newblock Analyzing spatial user behavior in computer games using geographic
  information systems.
\newblock In {\em Proceedings of the 13th International {MindTrek} Conference},
  pp. 182--189. {ACM}, 2009. \href{https://doi.org/10.1145/1621841.1621875}
{doi: {{%
10\hspace{.1pt}\discretionary{.}{%
}{.}\hspace{.4pt}1145\discretionary{/}{%
}{/}1621841\hspace{.1pt}\discretionary{.}{%
}{.}\hspace{.4pt}1621875}}}


\bibitem{Drachen2009-ys}
A.~Drachen and A.~Canossa.
\newblock Towards gameplay analysis via gameplay metrics.
\newblock In {\em Proceedings of the 13th International {MindTrek} Conference},
  pp. 202--209. {ACM}, 2009. \href{https://doi.org/10.1145/1621841.1621878}
{doi: {{%
10\hspace{.1pt}\discretionary{.}{%
}{.}\hspace{.4pt}1145\discretionary{/}{%
}{/}1621841\hspace{.1pt}\discretionary{.}{%
}{.}\hspace{.4pt}1621878}}}


\bibitem{Drachen2009-bc}
A.~Drachen, A.~Canossa, and G.~N. Yannakakis.
\newblock Player modeling using self-organization in tomb raider: Underworld.
\newblock In {\em Proceedings of the {IEEE} Symposium on Computational
  Intelligence and Games}, pp. 1--8, 2009.
  \href{https://doi.org/10.1109/CIG.2009.5286500}
{doi: {{%
10\hspace{.1pt}\discretionary{.}{%
}{.}\hspace{.4pt}1109\discretionary{/}{%
}{/}CIG\hspace{.1pt}\discretionary{.}{%
}{.}\hspace{.4pt}2009\hspace{.1pt}\discretionary{.}{%
}{.}\hspace{.4pt}5286500}}}


\bibitem{dutta2019via}
A.~Dutta and A.~Zisserman.
\newblock The via annotation software for images, audio and video.
\newblock In {\em Proceedings of the 27th {ACM} International Conference on
  Multimedia}, pp. 2276--2279, 2019.
  \href{https://doi.org/10.1145/3343031.3350535}
{doi: {{%
10\hspace{.1pt}\discretionary{.}{%
}{.}\hspace{.4pt}1145\discretionary{/}{%
}{/}3343031\hspace{.1pt}\discretionary{.}{%
}{.}\hspace{.4pt}3350535}}}


\bibitem{Fanaee-T2016-sy}
H.~Fanaee-T and J.~Gama.
\newblock Tensor-based anomaly detection: An interdisciplinary survey.
\newblock {\em Knowledge-Based Systems}, 98:130--147, 2016.
  \href{https://doi.org/10.1016/j.knosys.2016.01.027}
{doi: {{%
10\hspace{.1pt}\discretionary{.}{%
}{.}\hspace{.4pt}1016\discretionary{/}{%
}{/}j\hspace{.1pt}\discretionary{.}{%
}{.}\hspace{.4pt}knosys\hspace{.1pt}\discretionary{.}{%
}{.}\hspace{.4pt}2016\hspace{.1pt}\discretionary{.}{%
}{.}\hspace{.4pt}01\hspace{.1pt}\discretionary{.}{%
}{.}\hspace{.4pt}027}}}


\bibitem{Foo2004-oi}
C.~Y. Foo and E.~M.~I. Koivisto.
\newblock Defining grief play in {MMORPGs}: player and developer perceptions.
\newblock In {\em Proceedings of the {ACM} {SIGCHI} International Conference on
  Advances in Computer Entertainment Technology}, pp. 245--250, 2004.
  \href{https://doi.org/10.1145/1067343.1067375}
{doi: {{%
10\hspace{.1pt}\discretionary{.}{%
}{.}\hspace{.4pt}1145\discretionary{/}{%
}{/}1067343\hspace{.1pt}\discretionary{.}{%
}{.}\hspace{.4pt}1067375}}}


\bibitem{Geiger2020-ed}
A.~Geiger, D.~Liu, S.~Alnegheimish, A.~Cuesta-Infante, and K.~Veeramachaneni.
\newblock {TadGAN}: Time series anomaly detection using generative adversarial
  networks.
\newblock In {\em {IEEE} International Conference on Big Data}, pp. 33--43,
  2020. \href{https://doi.org/10.1109/BigData50022.2020.9378139}
{doi: {{%
10\hspace{.1pt}\discretionary{.}{%
}{.}\hspace{.4pt}1109\discretionary{/}{%
}{/}BigData50022\hspace{.1pt}\discretionary{.}{%
}{.}\hspace{.4pt}2020\hspace{.1pt}\discretionary{.}{%
}{.}\hspace{.4pt}9378139}}}


\bibitem{Grandprey-Shores2014-ra}
K.~Grandprey-Shores, Y.~He, K.~L. Swanenburg, R.~Kraut, and J.~Riedl.
\newblock The identification of deviance and its impact on retention in a
  multiplayer game.
\newblock In {\em Proceedings of the 17th {ACM} Conference on Computer
  Supported Cooperative Work}, pp. 1356--1365, 2014.
  \href{https://doi.org/10.1145/2531602.2531724}
{doi: {{%
10\hspace{.1pt}\discretionary{.}{%
}{.}\hspace{.4pt}1145\discretionary{/}{%
}{/}2531602\hspace{.1pt}\discretionary{.}{%
}{.}\hspace{.4pt}2531724}}}


\bibitem{Guo2019-pv}
S.~Guo, Z.~Jin, Q.~Chen, D.~Gotz, H.~Zha, and N.~Cao.
\newblock Visual anomaly detection in event sequence data.
\newblock In {\em {IEEE} International Conference on Big Data}, pp. 1125--1130,
  2019. \href{https://doi.org/10.1109/BigData47090.2019.9005687}
{doi: {{%
10\hspace{.1pt}\discretionary{.}{%
}{.}\hspace{.4pt}1109\discretionary{/}{%
}{/}BigData47090\hspace{.1pt}\discretionary{.}{%
}{.}\hspace{.4pt}2019\hspace{.1pt}\discretionary{.}{%
}{.}\hspace{.4pt}9005687}}}


\bibitem{Hodge2004-ta}
V.~Hodge and J.~Austin.
\newblock A survey of outlier detection methodologies.
\newblock {\em Artificial Intelligence Review}, 22(2):85--126, 2004.
  \href{https://doi.org/10.1023/B:AIRE.0000045502.10941.a9}
{doi: {{%
10\hspace{.1pt}\discretionary{.}{%
}{.}\hspace{.4pt}1023\discretionary{/}{%
}{/}B\discretionary{:}{%
}{:}AIRE\hspace{.1pt}\discretionary{.}{%
}{.}\hspace{.4pt}0000045502\hspace{.1pt}\discretionary{.}{%
}{.}\hspace{.4pt}10941\hspace{.1pt}\discretionary{.}{%
}{.}\hspace{.4pt}a9}}}


\bibitem{Javvaji2020-eu}
N.~Javvaji, C.~Harteveld, and M.~Seif El-Nasr.
\newblock Understanding player patterns by combining knowledge-based data
  abstraction with interactive visualization.
\newblock In {\em Proceedings of the Annual Symposium on {Computer-Human}
  Interaction in Play}, pp. 254--266. {ACM}, 2020.
  \href{https://doi.org/10.1145/3410404.3414257}
{doi: {{%
10\hspace{.1pt}\discretionary{.}{%
}{.}\hspace{.4pt}1145\discretionary{/}{%
}{/}3410404\hspace{.1pt}\discretionary{.}{%
}{.}\hspace{.4pt}3414257}}}


\bibitem{Kokkinakis2020-uo}
A.~V. Kokkinakis, S.~Demediuk, I.~Nölle, O.~Olarewaju, S.~Patra, J.~Robertson,
  P.~York, A.~P. Pedrassoli~Chitayat, A.~Coates, D.~Slawson, P.~Hughes,
  N.~Hardie, B.~Kirman, J.~Hook, A.~Drachen, M.~F. Ursu, and F.~Block.
\newblock {DAX}: Data-driven audience experiences in esports.
\newblock In {\em Proceedings of the {ACM} International Conference on
  Interactive Media Experiences}, pp. 94--105, 2020.
  \href{https://doi.org/10.1145/3391614.3393659}
{doi: {{%
10\hspace{.1pt}\discretionary{.}{%
}{.}\hspace{.4pt}1145\discretionary{/}{%
}{/}3391614\hspace{.1pt}\discretionary{.}{%
}{.}\hspace{.4pt}3393659}}}


\bibitem{Kordyaka2022-zi}
B.~Kordyaka, J.~Krath, S.~Park, H.~Wesseloh, and S.~Laato.
\newblock Understanding toxicity in multiplayer online games: The roles of
  national culture and demographic variables.
\newblock In {\em Proceedings of the Annual Hawaii International Conference on
  System Sciences}, 2022. \href{https://doi.org/10.24251/hicss.2022.359}
{doi: {{%
10\hspace{.1pt}\discretionary{.}{%
}{.}\hspace{.4pt}24251\discretionary{/}{%
}{/}hicss\hspace{.1pt}\discretionary{.}{%
}{.}\hspace{.4pt}2022\hspace{.1pt}\discretionary{.}{%
}{.}\hspace{.4pt}359}}}


\bibitem{Kou2020-eh}
Y.~Kou.
\newblock Toxic behaviors in team-based competitive gaming: The case of league
  of legends.
\newblock In {\em Proceedings of the Annual Symposium on {Computer-Human}
  Interaction in Play}, pp. 81--92. {ACM}, 2020.
  \href{https://doi.org/10.1145/3410404.3414243}
{doi: {{%
10\hspace{.1pt}\discretionary{.}{%
}{.}\hspace{.4pt}1145\discretionary{/}{%
}{/}3410404\hspace{.1pt}\discretionary{.}{%
}{.}\hspace{.4pt}3414243}}}


\bibitem{Kou2017-nx}
Y.~Kou, X.~Gui, S.~Zhang, and B.~Nardi.
\newblock Managing disruptive behavior through non-hierarchical governance:
  Crowdsourcing in league of legends and weibo.
\newblock {\em Proceedings of the ACM on Human-Computer Interaction}, 1:1--17,
  2017. \href{https://doi.org/10.1145/3134697}
{doi: {{%
10\hspace{.1pt}\discretionary{.}{%
}{.}\hspace{.4pt}1145\discretionary{/}{%
}{/}3134697}}}


\bibitem{Kuan2017-ye}
Y.-T. Kuan, Y.-S. Wang, and J.-H. Chuang.
\newblock Visualizing real-time strategy games: The example of {StarCraft}
  {II}.
\newblock In {\em {IEEE} Conference on Visual Analytics Science and
  Technology}, pp. 71--80, 2017.
  \href{https://doi.org/10.1109/VAST.2017.8585594}
{doi: {{%
10\hspace{.1pt}\discretionary{.}{%
}{.}\hspace{.4pt}1109\discretionary{/}{%
}{/}VAST\hspace{.1pt}\discretionary{.}{%
}{.}\hspace{.4pt}2017\hspace{.1pt}\discretionary{.}{%
}{.}\hspace{.4pt}8585594}}}


\bibitem{kucher2017active}
K.~Kucher, C.~Paradis, M.~Sahlgren, and A.~Kerren.
\newblock Active learning and visual analytics for stance classification with
  alva.
\newblock {\em ACM Transactions on Interactive Intelligent Systems},
  7(3):1--31, 2017. \href{https://doi.org/10.1145/3132169}
{doi: {{%
10\hspace{.1pt}\discretionary{.}{%
}{.}\hspace{.4pt}1145\discretionary{/}{%
}{/}3132169}}}


\bibitem{kwak2015exploring}
H.~Kwak, J.~Blackburn, and S.~Han.
\newblock Exploring cyberbullying and other toxic behavior in team competition
  online games.
\newblock In {\em Proceedings of the SIGCHI Conference on Human Factors in
  Computing Systems}, pp. 3739--3748. {ACM}, 2015.
  \href{https://doi.org/10.1145/2702123.2702529}
{doi: {{%
10\hspace{.1pt}\discretionary{.}{%
}{.}\hspace{.4pt}1145\discretionary{/}{%
}{/}2702123\hspace{.1pt}\discretionary{.}{%
}{.}\hspace{.4pt}2702529}}}


\bibitem{Li2018-kb}
Q.~Li, Z.~Wu, P.~Xu, H.~Qu, and X.~Ma.
\newblock A multi-phased co-design of an interactive analytics system for moba
  game occurrences.
\newblock In {\em Proceedings of the Designing Interactive Systems Conference},
  pp. 1321--1332. {ACM}, 2018. \href{https://doi.org/10.1145/3196709.3196771}
{doi: {{%
10\hspace{.1pt}\discretionary{.}{%
}{.}\hspace{.4pt}1145\discretionary{/}{%
}{/}3196709\hspace{.1pt}\discretionary{.}{%
}{.}\hspace{.4pt}3196771}}}


\bibitem{Li2017-dn}
Q.~Li, P.~Xu, Y.~Y. Chan, Y.~Wang, Z.~Wang, H.~Qu, and X.~Ma.
\newblock A visual analytics approach for understanding reasons behind
  snowballing and comeback in {MOBA} games.
\newblock {\em IEEE Transactions on Visualization and Computer Graphics},
  23(1):211--220, 2017. \href{https://doi.org/10.1109/TVCG.2016.2598415}
{doi: {{%
10\hspace{.1pt}\discretionary{.}{%
}{.}\hspace{.4pt}1109\discretionary{/}{%
}{/}TVCG\hspace{.1pt}\discretionary{.}{%
}{.}\hspace{.4pt}2016\hspace{.1pt}\discretionary{.}{%
}{.}\hspace{.4pt}2598415}}}


\bibitem{liao2016visualization}
H.~Liao, L.~Chen, Y.~Song, and H.~Ming.
\newblock Visualization-based active learning for video annotation.
\newblock {\em IEEE Transactions on Multimedia}, 18(11):2196--2205, 2016.
  \href{https://doi.org/10.1109/TMM.2016.2614227}
{doi: {{%
10\hspace{.1pt}\discretionary{.}{%
}{.}\hspace{.4pt}1109\discretionary{/}{%
}{/}TMM\hspace{.1pt}\discretionary{.}{%
}{.}\hspace{.4pt}2016\hspace{.1pt}\discretionary{.}{%
}{.}\hspace{.4pt}2614227}}}


\bibitem{Liao2010-oo}
Z.~Liao, Y.~Yu, and B.~Chen.
\newblock Anomaly detection in {GPS} data based on visual analytics.
\newblock In {\em {IEEE} Conference on Visual Analytics Science and
  Technology}, pp. 51--58, 2010.
  \href{https://doi.org/10.1109/VAST.2010.5652467}
{doi: {{%
10\hspace{.1pt}\discretionary{.}{%
}{.}\hspace{.4pt}1109\discretionary{/}{%
}{/}VAST\hspace{.1pt}\discretionary{.}{%
}{.}\hspace{.4pt}2010\hspace{.1pt}\discretionary{.}{%
}{.}\hspace{.4pt}5652467}}}


\bibitem{Lin2018-er}
C.~Lin, Q.~Zhu, S.~Guo, Z.~Jin, Y.-R. Lin, and N.~Cao.
\newblock Anomaly detection in spatiotemporal data via regularized non-negative
  tensor analysis.
\newblock {\em Data Mining and Knowledge Discovery}, 32(4):1056--1073, 2018.
  \href{https://doi.org/10.1007/s10618-018-0560-3}
{doi: {{%
10\hspace{.1pt}\discretionary{.}{%
}{.}\hspace{.4pt}1007\discretionary{/}{%
}{/}s10618\discretionary{%
}{-}{-}018\discretionary{%
}{-}{-}0560\discretionary{%
}{-}{-}3}}}


\bibitem{Liu2021-nb}
D.~Liu, S.~Alnegheimish, A.~Zytek, and {others}.
\newblock {MTV}: Visual analytics for detecting, investigating, and annotating
  anomalies in multivariate time series.
\newblock {\em Proceedings of the ACM on Human-Computer Interaction}, 6:1--30,
  2022. \href{https://doi.org/10.1145/3512950}
{doi: {{%
10\hspace{.1pt}\discretionary{.}{%
}{.}\hspace{.4pt}1145\discretionary{/}{%
}{/}3512950}}}


\bibitem{Lu2019-zf}
J.~Lu, X.~Xie, J.~Lan, T.-Q. Peng, W.~Chen, and Y.~Wu.
\newblock Visual analytics of dynamic interplay between behaviors in {MMORPGs}.
\newblock In {\em {IEEE} Pacific Visualization Symposium}, pp. 112--121, 2019.
  \href{https://doi.org/10.1109/PacificVis.2019.00021}
{doi: {{%
10\hspace{.1pt}\discretionary{.}{%
}{.}\hspace{.4pt}1109\discretionary{/}{%
}{/}PacificVis\hspace{.1pt}\discretionary{.}{%
}{.}\hspace{.4pt}2019\hspace{.1pt}\discretionary{.}{%
}{.}\hspace{.4pt}00021}}}


\bibitem{mcinnes2018umap}
L.~McInnes, J.~Healy, and J.~Melville.
\newblock Umap: Uniform manifold approximation and projection for dimension
  reduction.
\newblock {\em arXiv preprint arXiv:1802.03426}, 2018.

\bibitem{McKenna2016-py}
S.~McKenna, D.~Staheli, C.~Fulcher, and M.~Meyer.
\newblock {BubbleNet}: A cyber security dashboard for visualizing patterns.
\newblock {\em Computer Graphics Forum}, 35(3):281--290, 2016.
  \href{https://doi.org/10.1111/cgf.12904}
{doi: {{%
10\hspace{.1pt}\discretionary{.}{%
}{.}\hspace{.4pt}1111\discretionary{/}{%
}{/}cgf\hspace{.1pt}\discretionary{.}{%
}{.}\hspace{.4pt}12904}}}


\bibitem{Medler2011-fd}
B.~Medler, M.~John, and J.~Lane.
\newblock Data cracker: Developing a visual game analytic tool for analyzing
  online gameplay.
\newblock In {\em Proceedings of the {SIGCHI} Conference on Human Factors in
  Computing Systems}, pp. 2365--2374. {ACM}, 2011.
  \href{https://doi.org/10.1145/1978942.1979288}
{doi: {{%
10\hspace{.1pt}\discretionary{.}{%
}{.}\hspace{.4pt}1145\discretionary{/}{%
}{/}1978942\hspace{.1pt}\discretionary{.}{%
}{.}\hspace{.4pt}1979288}}}


\bibitem{Milam2010-dw}
D.~Milam and M.~S. El~Nasr.
\newblock Design patterns to guide player movement in {3D} games.
\newblock In {\em Proceedings of the 5th {ACM} {SIGGRAPH} Symposium on Video
  Games}, pp. 37--42, 2010. \href{https://doi.org/10.1145/1836135.1836141}
{doi: {{%
10\hspace{.1pt}\discretionary{.}{%
}{.}\hspace{.4pt}1145\discretionary{/}{%
}{/}1836135\hspace{.1pt}\discretionary{.}{%
}{.}\hspace{.4pt}1836141}}}


\bibitem{Mu2019-yq}
X.~Mu, K.~Xu, Q.~Chen, F.~Du, Y.~Wang, and H.~Qu.
\newblock {MOOCad}: Visual analysis of anomalous learning activities in massive
  open online courses.
\newblock In {\em Proceedings of the 21st Eurographics Conference on
  Visualization}, pp. 91--95, 2019.
  \href{https://doi.org/10.2312/evs20191176/091-095}
{doi: {{%
10\hspace{.1pt}\discretionary{.}{%
}{.}\hspace{.4pt}2312\discretionary{/}{%
}{/}evs20191176\discretionary{/}{%
}{/}091\discretionary{%
}{-}{-}095}}}


\bibitem{Martens2015-qj}
M.~Märtens, S.~Shen, A.~Iosup, and F.~Kuipers.
\newblock Toxicity detection in multiplayer online games.
\newblock In {\em International Workshop on Network and Systems Support for
  Games}, pp. 1--6, 2015. \href{https://doi.org/10.1109/NetGames.2015.7382991}
{doi: {{%
10\hspace{.1pt}\discretionary{.}{%
}{.}\hspace{.4pt}1109\discretionary{/}{%
}{/}NetGames\hspace{.1pt}\discretionary{.}{%
}{.}\hspace{.4pt}2015\hspace{.1pt}\discretionary{.}{%
}{.}\hspace{.4pt}7382991}}}


\bibitem{paiva2014approach}
J.~G.~S. Paiva, W.~R. Schwartz, H.~Pedrini, and R.~Minghim.
\newblock An approach to supporting incremental visual data classification.
\newblock {\em IEEE Transactions on Visualization and Computer Graphics},
  21(1):4--17, 2014. \href{https://doi.org/10.1109/TVCG.2014.2331979}
{doi: {{%
10\hspace{.1pt}\discretionary{.}{%
}{.}\hspace{.4pt}1109\discretionary{/}{%
}{/}TVCG\hspace{.1pt}\discretionary{.}{%
}{.}\hspace{.4pt}2014\hspace{.1pt}\discretionary{.}{%
}{.}\hspace{.4pt}2331979}}}


\bibitem{Pimentel2020-eh}
T.~Pimentel, M.~Monteiro, A.~Veloso, and N.~Ziviani.
\newblock Deep active learning for anomaly detection.
\newblock In {\em International Joint Conference on Neural Networks}, pp. 1--8,
  2020. \href{https://doi.org/10.1109/IJCNN48605.2020.9206769}
{doi: {{%
10\hspace{.1pt}\discretionary{.}{%
}{.}\hspace{.4pt}1109\discretionary{/}{%
}{/}IJCNN48605\hspace{.1pt}\discretionary{.}{%
}{.}\hspace{.4pt}2020\hspace{.1pt}\discretionary{.}{%
}{.}\hspace{.4pt}9206769}}}


\bibitem{pohjanen2018report}
A.~Pohjanen.
\newblock Report please! {A} survey on players' perceptions towards the tools
  for fighting toxic behavior in competitive online multiplayer video games.
\newblock 2018.

\bibitem{Rousseeuw2005-hh}
P.~J. Rousseeuw and A.~M. Leroy.
\newblock {\em Robust Regression and Outlier Detection}.
\newblock John Wiley \& Sons, 2005. \href{https://doi.org/10.1002/0471725382}
{doi: {{%
10\hspace{.1pt}\discretionary{.}{%
}{.}\hspace{.4pt}1002\discretionary{/}{%
}{/}0471725382}}}


\bibitem{Russo2020-xi}
S.~Russo, M.~Lürig, W.~Hao, B.~Matthews, and K.~Villez.
\newblock Active learning for anomaly detection in environmental data.
\newblock {\em Environmental Modelling \& Software}, 134:104869, 2020.
  \href{https://doi.org/10.1016/j.envsoft.2020.104869}
{doi: {{%
10\hspace{.1pt}\discretionary{.}{%
}{.}\hspace{.4pt}1016\discretionary{/}{%
}{/}j\hspace{.1pt}\discretionary{.}{%
}{.}\hspace{.4pt}envsoft\hspace{.1pt}\discretionary{.}{%
}{.}\hspace{.4pt}2020\hspace{.1pt}\discretionary{.}{%
}{.}\hspace{.4pt}104869}}}


\bibitem{settles2009active}
B.~Settles.
\newblock Active learning literature survey.
\newblock 2009.

\bibitem{Smiti2013-bb}
A.~Smiti and Z.~Eloudi.
\newblock Soft {DBSCAN}: Improving {DBSCAN} clustering method using fuzzy set
  theory.
\newblock In {\em Proceedings of the 6th International Conference on Human
  System Interactions}, pp. 380--385, 2013.
  \href{https://doi.org/10.1109/HSI.2013.6577851}
{doi: {{%
10\hspace{.1pt}\discretionary{.}{%
}{.}\hspace{.4pt}1109\discretionary{/}{%
}{/}HSI\hspace{.1pt}\discretionary{.}{%
}{.}\hspace{.4pt}2013\hspace{.1pt}\discretionary{.}{%
}{.}\hspace{.4pt}6577851}}}


\bibitem{Smiti2012-cf}
A.~Smiti and Z.~Elouedi.
\newblock {DBSCAN-GM}: An improved clustering method based on gaussian means
  and {DBSCAN} techniques.
\newblock In {\em {IEEE} 16th International Conference on Intelligent
  Engineering Systems}, pp. 573--578, 2012.
  \href{https://doi.org/10.1109/INES.2012.6249802}
{doi: {{%
10\hspace{.1pt}\discretionary{.}{%
}{.}\hspace{.4pt}1109\discretionary{/}{%
}{/}INES\hspace{.1pt}\discretionary{.}{%
}{.}\hspace{.4pt}2012\hspace{.1pt}\discretionary{.}{%
}{.}\hspace{.4pt}6249802}}}


\bibitem{suh2007semi}
B.~Suh and B.~B. Bederson.
\newblock Semi-automatic photo annotation strategies using event based
  clustering and clothing based person recognition.
\newblock {\em Interacting with Computers}, 19(4):524--544, 2007.
  \href{https://doi.org/10.1016/j.intcom.2007.02.002}
{doi: {{%
10\hspace{.1pt}\discretionary{.}{%
}{.}\hspace{.4pt}1016\discretionary{/}{%
}{/}j\hspace{.1pt}\discretionary{.}{%
}{.}\hspace{.4pt}intcom\hspace{.1pt}\discretionary{.}{%
}{.}\hspace{.4pt}2007\hspace{.1pt}\discretionary{.}{%
}{.}\hspace{.4pt}02\hspace{.1pt}\discretionary{.}{%
}{.}\hspace{.4pt}002}}}


\bibitem{tang2013towards}
J.~Tang, Q.~Chen, M.~Wang, S.~Yan, T.-S. Chua, and R.~Jain.
\newblock Towards optimizing human labeling for interactive image tagging.
\newblock {\em ACM Transactions on Multimedia Computing, Communications, and
  Applications}, 9(4):1--18, 2013.
  \href{https://doi.org/10.1145/2501643.2501651}
{doi: {{%
10\hspace{.1pt}\discretionary{.}{%
}{.}\hspace{.4pt}1145\discretionary{/}{%
}{/}2501643\hspace{.1pt}\discretionary{.}{%
}{.}\hspace{.4pt}2501651}}}


\bibitem{Thawonmas2008-ny}
R.~Thawonmas and K.~Iizuka.
\newblock Visualization of online-game players based on their action behaviors.
\newblock {\em International Journal of Computer Games Technology}, 2008:1--9,
  2008. \href{https://doi.org/10.1155/2008/906931}
{doi: {{%
10\hspace{.1pt}\discretionary{.}{%
}{.}\hspace{.4pt}1155\discretionary{/}{%
}{/}2008\discretionary{/}{%
}{/}906931}}}


\bibitem{Thawonmas2006-rd}
R.~Thawonmas, M.~Kurashige, K.~Iizuka, and M.~M. Kantardzic.
\newblock Clustering of online game users based on their trails using
  self-organizing map.
\newblock In {\em Proceedings of the International Conference on Entertainment
  Computing}, vol. 4161, pp. 366--369. Springer, 2006.
  \href{https://doi.org/10.1007/11872320_51}
{doi: {{%
10\hspace{.1pt}\discretionary{.}{%
}{.}\hspace{.4pt}1007\discretionary{/}{%
}{/}11872320\_51}}}


\bibitem{torralba2010labelme}
A.~Torralba, B.~C. Russell, and J.~Yuen.
\newblock Labelme: Online image annotation and applications.
\newblock {\em Proceedings of the IEEE}, 98(8):1467--1484, 2010.
  \href{https://doi.org/10.1109/JPROC.2010.2050290}
{doi: {{%
10\hspace{.1pt}\discretionary{.}{%
}{.}\hspace{.4pt}1109\discretionary{/}{%
}{/}JPROC\hspace{.1pt}\discretionary{.}{%
}{.}\hspace{.4pt}2010\hspace{.1pt}\discretionary{.}{%
}{.}\hspace{.4pt}2050290}}}


\bibitem{Vaz2021-ka}
A.~Vaz, A.~Batista, F.~Silva, L.~M. Costa, and G.~Xexéo.
\newblock A low-code approach to identify toxicity in {MOBA} games.
\newblock In {\em Proceedings of the 20th Brazilian Symposium on Games and
  Digital Entertainment}, pp. 305--308. SBC, 2021.
  \href{https://doi.org/10.5753/sbgames_estendido.2021.19657}
{doi: {{%
10\hspace{.1pt}\discretionary{.}{%
}{.}\hspace{.4pt}5753\discretionary{/}{%
}{/}sbgames\_estendido\hspace{.1pt}\discretionary{.}{%
}{.}\hspace{.4pt}2021\hspace{.1pt}\discretionary{.}{%
}{.}\hspace{.4pt}19657}}}


\bibitem{Wallner2011-fe}
G.~Wallner and S.~Kriglstein.
\newblock Design and evaluation of the educational game {DOGeometry}: {A} case
  study.
\newblock In {\em Proceedings of the 8th International Conference on Advances
  in Computer Entertainment Technology}, pp. 1--8. {ACM}, 2011.
  \href{https://doi.org/10.1145/2071423.2071441}
{doi: {{%
10\hspace{.1pt}\discretionary{.}{%
}{.}\hspace{.4pt}1145\discretionary{/}{%
}{/}2071423\hspace{.1pt}\discretionary{.}{%
}{.}\hspace{.4pt}2071441}}}


\bibitem{Wallner2013-um}
G.~Wallner and S.~Kriglstein.
\newblock Visualization-based analysis of gameplay data – {A} review of
  literature.
\newblock {\em Entertainment Computing}, 4(3):143--155, 2013.
  \href{https://doi.org/10.1016/j.entcom.2013.02.002}
{doi: {{%
10\hspace{.1pt}\discretionary{.}{%
}{.}\hspace{.4pt}1016\discretionary{/}{%
}{/}j\hspace{.1pt}\discretionary{.}{%
}{.}\hspace{.4pt}entcom\hspace{.1pt}\discretionary{.}{%
}{.}\hspace{.4pt}2013\hspace{.1pt}\discretionary{.}{%
}{.}\hspace{.4pt}02\hspace{.1pt}\discretionary{.}{%
}{.}\hspace{.4pt}002}}}


\bibitem{Wallner2021-fy}
G.~Wallner, M.~van Wijland, R.~Bernhaupt, and S.~Kriglstein.
\newblock What players want: Information needs of players on post-game
  visualizations.
\newblock In {\em Proceedings of the SIGCHI Conference on Human Factors in
  Computing Systems}, pp. 1--13. {ACM}, 2021.
  \href{https://doi.org/10.1145/3411764.3445174}
{doi: {{%
10\hspace{.1pt}\discretionary{.}{%
}{.}\hspace{.4pt}1145\discretionary{/}{%
}{/}3411764\hspace{.1pt}\discretionary{.}{%
}{.}\hspace{.4pt}3445174}}}


\bibitem{Wong2022-im}
L.~Wong, D.~Liu, L.~Berti-Equille, S.~Alnegheimish, and K.~Veeramachaneni.
\newblock {AER}: {Auto-Encoder} with regression for time series anomaly
  detection.
\newblock In {\em {IEEE} International Conference on Big Data}, pp. 1152--1161,
  2022. \href{https://doi.org/10.1109/BigData55660.2022.10020857}
{doi: {{%
10\hspace{.1pt}\discretionary{.}{%
}{.}\hspace{.4pt}1109\discretionary{/}{%
}{/}BigData55660\hspace{.1pt}\discretionary{.}{%
}{.}\hspace{.4pt}2022\hspace{.1pt}\discretionary{.}{%
}{.}\hspace{.4pt}10020857}}}


\bibitem{Xie2022-ku}
L.~Xie, Z.~Wu, P.~Xu, W.~Li, X.~Ma, and Q.~Li.
\newblock {RoleSeer}: Understanding informal social role changes in {MMORPGs}
  via visual analytics.
\newblock In {\em Proceedings of the SIGCHI Conference on Human Factors in
  Computing Systems}, pp. 1--17. {ACM}, 2022.
  \href{https://doi.org/10.1145/3491102.3517712}
{doi: {{%
10\hspace{.1pt}\discretionary{.}{%
}{.}\hspace{.4pt}1145\discretionary{/}{%
}{/}3491102\hspace{.1pt}\discretionary{.}{%
}{.}\hspace{.4pt}3517712}}}


\bibitem{Xu2020-mu}
K.~Xu, Y.~Wang, L.~Yang, Y.~Wang, B.~Qiao, S.~Qin, Y.~Xu, H.~Zhang, and H.~Qu.
\newblock {CloudDet}: Interactive visual analysis of anomalous performances in
  cloud computing systems.
\newblock {\em IEEE Transactions on Visualization and Computer Graphics},
  26(1):1107--1117, 2020. \href{https://doi.org/10.1109/TVCG.2019.2934613}
{doi: {{%
10\hspace{.1pt}\discretionary{.}{%
}{.}\hspace{.4pt}1109\discretionary{/}{%
}{/}TVCG\hspace{.1pt}\discretionary{.}{%
}{.}\hspace{.4pt}2019\hspace{.1pt}\discretionary{.}{%
}{.}\hspace{.4pt}2934613}}}


\bibitem{Xu2018-ll}
K.~Xu, M.~Xia, X.~Mu, Y.~Wang, and {others}.
\newblock {EnsembleLens}: Ensemble-based visual exploration of anomaly
  detection algorithms with multidimensional data.
\newblock {\em IEEE Transactions on Visualization and Computer Graphics},
  25(1):109--119, 2019. \href{https://doi.org/10.1109/TVCG.2018.2864825}
{doi: {{%
10\hspace{.1pt}\discretionary{.}{%
}{.}\hspace{.4pt}1109\discretionary{/}{%
}{/}TVCG\hspace{.1pt}\discretionary{.}{%
}{.}\hspace{.4pt}2018\hspace{.1pt}\discretionary{.}{%
}{.}\hspace{.4pt}2864825}}}


\bibitem{zhang2021mi3}
Y.~Zhang, B.~Coecke, and M.~Chen.
\newblock Mi3: Machine-initiated intelligent interaction for interactive
  classification and data reconstruction.
\newblock {\em ACM Transactions on Interactive Intelligent Systems}, 11:1--34,
  2021. \href{https://doi.org/10.1145/3412848}
{doi: {{%
10\hspace{.1pt}\discretionary{.}{%
}{.}\hspace{.4pt}1145\discretionary{/}{%
}{/}3412848}}}


\bibitem{zhang2022onelabeler}
Y.~Zhang, Y.~Wang, H.~Zhang, B.~Zhu, S.~Chen, and D.~Zhang.
\newblock Onelabeler: A flexible system for building data labeling tools.
\newblock In {\em Proceedings of the SIGCHI Conference on Human Factors in
  Computing Systems}, pp. 1--22. {ACM}, 2022.
  \href{https://doi.org/10.1145/3491102.3517612}
{doi: {{%
10\hspace{.1pt}\discretionary{.}{%
}{.}\hspace{.4pt}1145\discretionary{/}{%
}{/}3491102\hspace{.1pt}\discretionary{.}{%
}{.}\hspace{.4pt}3517612}}}


\bibitem{Zhao2014-xv}
J.~Zhao, N.~Cao, Z.~Wen, Y.~Song, Y.~Lin, and C.~Collins.
\newblock \#{FluxFlow}: Visual analysis of anomalous information spreading on
  social media.
\newblock {\em IEEE Transactions on Visualization and Computer Graphics},
  20(12):1773--1782, 2014. \href{https://doi.org/10.1109/TVCG.2014.2346922}
{doi: {{%
10\hspace{.1pt}\discretionary{.}{%
}{.}\hspace{.4pt}1109\discretionary{/}{%
}{/}TVCG\hspace{.1pt}\discretionary{.}{%
}{.}\hspace{.4pt}2014\hspace{.1pt}\discretionary{.}{%
}{.}\hspace{.4pt}2346922}}}


\bibitem{Zhao2019-rl}
X.~Zhao, W.~Cui, Y.~Wu, H.~Zhang, H.~Qu, and D.~Zhang.
\newblock Oui! {Outlier} interpretation on multi‐dimensional data via visual
  analytics.
\newblock {\em Computer Graphics Forum}, 38(3):213--224, 2019.
  \href{https://doi.org/10.1111/cgf.13683}
{doi: {{%
10\hspace{.1pt}\discretionary{.}{%
}{.}\hspace{.4pt}1111\discretionary{/}{%
}{/}cgf\hspace{.1pt}\discretionary{.}{%
}{.}\hspace{.4pt}13683}}}


\bibitem{Zsila2022-ov}
{\'A}.~Zsila, R.~Shabahang, M.~S. Aruguete, and G.~Orosz.
\newblock Toxic behaviors in online multiplayer games: Prevalence, perception,
  risk factors of victimization, and psychological consequences.
\newblock {\em Aggressive Behavior}, 48(3):356--364, 2022.
  \href{https://doi.org/10.1002/ab.22023}
{doi: {{%
10\hspace{.1pt}\discretionary{.}{%
}{.}\hspace{.4pt}1002\discretionary{/}{%
}{/}ab\hspace{.1pt}\discretionary{.}{%
}{.}\hspace{.4pt}22023}}}


\end{thebibliography}

\appendix 


\section{Details of Data Abstraction}

The subsequent paragraphs furnish a comprehensive introduction to each type of data.

\par\noindent\textbf{Player information.} This type of data encompasses various details regarding a player's standing within the game, including their skill level as indicated by their proficiency with specific in-game heroes and their ranking or grade within the game's competitive system. Such data may serve as a valuable tool for users seeking to ascertain the player's level of expertise and competencies within the game's context.

\par\noindent\textbf{Match summary information.} This type of data encapsulates various performance statistics pertaining to an individual player's participation in a specific match, including metrics such as total kills, assists, and deaths. Additionally, this category may contain more overarching information such as the duration of the match, the total amount of damage inflicted and received, and the total economy of the player. Notably, the number of reports filed against a player during a given match can also serve as a crucial indicator of their performance. The numerical data contained within this category afford a preliminary insight into a player's in-game contributions, thereby enabling users to efficiently evaluate their performance within the broader context of their team.

\par\noindent\textbf{Multivariate event sequence.} This data type encompasses comprehensive records detailing every instance of a player's death, tower destruction, or killing of a dragon during gameplay, with associated information on the heroes or objects involved and the identity of the killer. The timing and location of these events are also logged. Given their significance within the game's mechanics, such events provide valuable insights into the overall progression of the match. By analyzing user behavior in conjunction with the information contained in this category, users may obtain a more nuanced understanding of the underlying motivations behind players' actions.

\par\noindent\textbf{Multivariate time series data.} Multivariate time series data refers to a collection of diverse statistics, such as economic indicators, kill counts, and damage dealt and received, which are captured at 20-second intervals during the entirety of a given match. This data is instrumental in providing a comprehensive view of the dynamic state of players during gameplay and can be leveraged to conduct meticulous and specific event abstractions.

\par\noindent\textbf{Player movement data.} Player movement data pertains to the positional information of in-game heroes that is logged at 10-second intervals. This data is critical in providing insights into the locations and movements of players throughout the duration of the match. Furthermore, this dataset encompasses the movement patterns of all ten players involved in the game. By analyzing and contrasting the trajectory of the focal player with that of the other players, it is possible to discern any anomalies or deviations in the player's movements.

\par\noindent\textbf{Algorithmic derived data.} Algorithmic derived data refers to a set of preliminary statistics that offer a bird's eye view of matches. These include metrics such as idle time, which measures the amount of time spent away from the keyboard, healthy recall, which tracks the number of times a player returns to the highland without experiencing any health point loss, and surrender times, which counts the number of surrender votes initiated during the game. This supplementary information aids in comprehending the overall gameplay behavior at a macro level. Additionally, this data exhibits a strong correlation with adverse player behaviors and can be incorporated in the analysis of said players.

\section{Details of Events Abstraction}

The subsequent paragraphs introduce each type of event.

\noindent\textbf{Hero killing.} The act of eliminating enemy heroes is possible for players by directing their own heroes to engage in combat. Upon successfully diminishing the opposing hero's health points, the targeted hero will be deemed slain.

\noindent\textbf{Death.} In-game mortality is a possibility for players, as their controlled hero may succumb to defeat, requiring a period of time for the hero to revive before resuming engagement with enemy forces.

\noindent\textbf{Assist in killing.} Players can also assist in defeating enemy heroes by contributing to their demise without necessarily landing the final blow.

\noindent\textbf{Poke.} The term ``poke'' refers to the act of causing harm to an opposing hero. This action is considered to have occurred irrespective of whether the player delivers the final blow resulting in the hero's death. Thus, any form of damage dealt to the enemy hero, regardless of the outcome, is considered a poke.

\noindent\textbf{Monster/dragon killing.} Players have the opportunity to engage in the killing of various monsters located in the jungle areas of the game map. These actions can result in the acquisition of additional economic resources, as well as special effects that enhance the performance of the player's hero. Specifically, the killing of the red or blue buff can confer temporary boosts to damage output or recovery rate, respectively. Furthermore, successfully killing dragons can result in significant benefits for the player's team, which may contribute to an ultimate victory.

\noindent\textbf{Minion killing.} Players have the option to eliminate enemy minions that spawn within the top, middle, and bottom lanes of the game map. These actions can result in the acquisition of economic resources. However, it is essential to note that players must successfully land the final blow on a soldier to receive the full economic benefit.

\noindent\textbf{Turret destruction.} Players have the ability to engage in the destruction of enemy turrets, as well as to advance ally soldiers toward the enemy base. Upon the successful elimination of all turrets within a specific lane, players gain the ability to directly attack the enemy base. The destruction of the enemy base represents the ultimate objective of the game, and achieving this goal results in a victory for the player's team.

\noindent\textbf{Inaction.} If players abstain from participating in any of the aforementioned actions, they are deemed to be inaction during the corresponding period.

\section{Wander Around, Intentionally Killed by Dragon, and Not Participate in Key Defense of Team Base}

E5, an experienced player in MOBA games, is willing to help identify and label high-level actors in the game. He is willing to use the system to explore the behaviors of high-level actors in MOBA games.

\begin{figure*}[h]
\centering
\includegraphics[width=\linewidth]{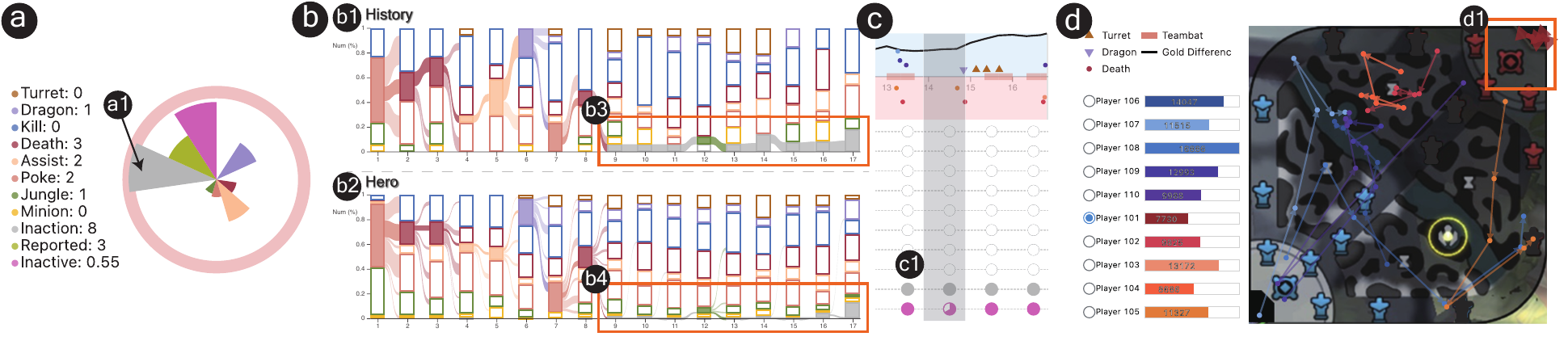}
\caption{This player is identified as an actor with the behavior of wandering around. (a) The glyph shows that the player has the highest number of inaction events. (b) In the history and hero mode of Progression View, the player has so many abnormal events during the $9^{th}$ to $17^{th}$ minutes. (c, d) A region is brushed, and the map chart demonstrates that the player is still staying in the highland.}
\label{fig:case_2_1}
\end{figure*}

\textbf{Wander around.} The number of focused players presented in the system is 30, as noticed in the Statistics Panel of the {\vone}. Then he adjusted the filters to focus on players with a high level of inaction, ranging from 3 to 8. He found one player with the highest number of inaction events (\autoref{fig:case_2_1}(a1)) and lasso-selected this player (\autoref{fig:case_2_1}(a)) (\textbf{R1}). He further contextualized the action patterns of this player in the history (\autoref{fig:case_2_1}(b1)) and hero mode (\autoref{fig:case_2_1}(b2)) of {\vtwo}. The results indicated that there are significant abnormal events between the $9^{th}$ and $17^{th}$ minutes (\autoref{fig:case_2_1}(b3, b4)), which consist of a large number of inaction events (\textbf{R2}). 
He believed that the particular player was likely an actor. To confirm his suspicion, he went through and brushed the region between the $9^{th}$ and $17^{th}$ minutes in the {\vfour}, but noticed that the player was not engaging in any active behaviors in the map chart (\autoref{fig:case_2_1}(c1, d1)). By analyzing the player's movements, he observed that the player was simply wandering around the highland or bottom lane (\textbf{R3}). He also clicked on the Profile button to check the player's profile information and discovered that the player had only been idle for 41 seconds. Despite the idle time not meeting the threshold, he categorized the player as an actor and suggested that the algorithm should be designed to identify players who have inaction events for more than eight minutes (\textbf{R4}).

\begin{figure*}[h]
\centering
\includegraphics[width=\linewidth]{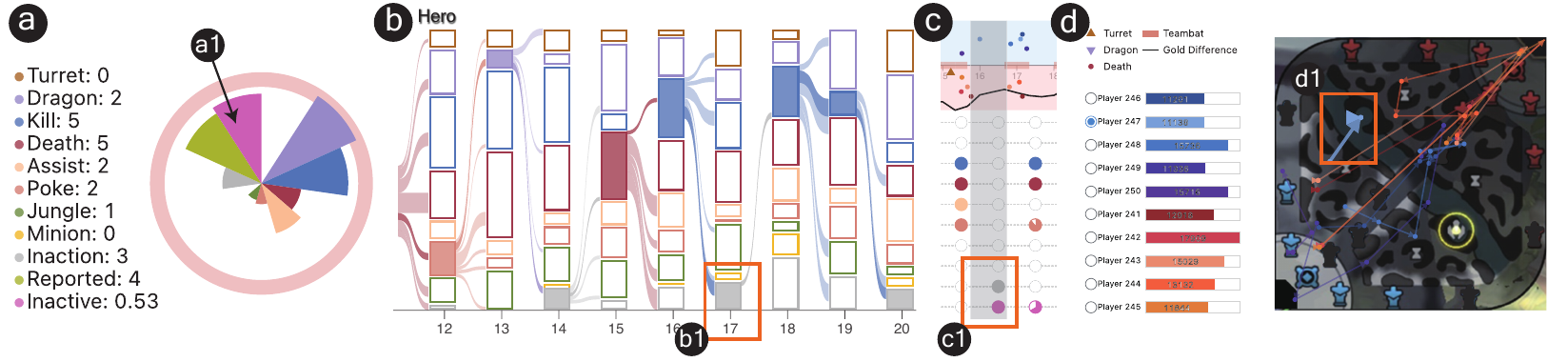}
\caption{The player is identified as an actor with the behavior of intentionally being killed by the dragon. (a) The glyph shows that the player has a large inactive percentage. (b) In the $17^{th}$ minute, the player has abnormal inaction events. (c, d) A region is brushed, and the map chart shows that the player is intentionally killed by the dragon.}
\label{fig:case_2_2}
\end{figure*}

\textbf{Intentionally killed by dragon.} E5 lasso-selected another player with a relatively large inactive percentage (\autoref{fig:case_2_2}(a1)) (\textbf{R1}) and checked the hero mode in {\vtwo} (\autoref{fig:case_2_2}(b)). He found that it is weird to have inaction events between the $16^{th}$ and $17^{th}$ minutes (\autoref{fig:case_2_2}(b1)), and the percentage of inaction events is quite small in this minute, indicating that other players typically do not have such events during this time period (\textbf{R2}). This raised E5's suspicion, prompting him to investigate further in the {\vfour}. E5 brushed those regions (\autoref{fig:case_2_2}(c1)) and found that the player attempted to kill the dragon but got killed (\autoref{fig:case_2_2}(d1)) (\textbf{R3}). Normally, the player will not be killed by the dragon, and other players should kill the dragon together. E5 found it suspicious and labeled it as an actor (\textbf{R4}). E5 then moved on to examine the next player.

\begin{figure*}[h]
\centering
\includegraphics[width=\linewidth]{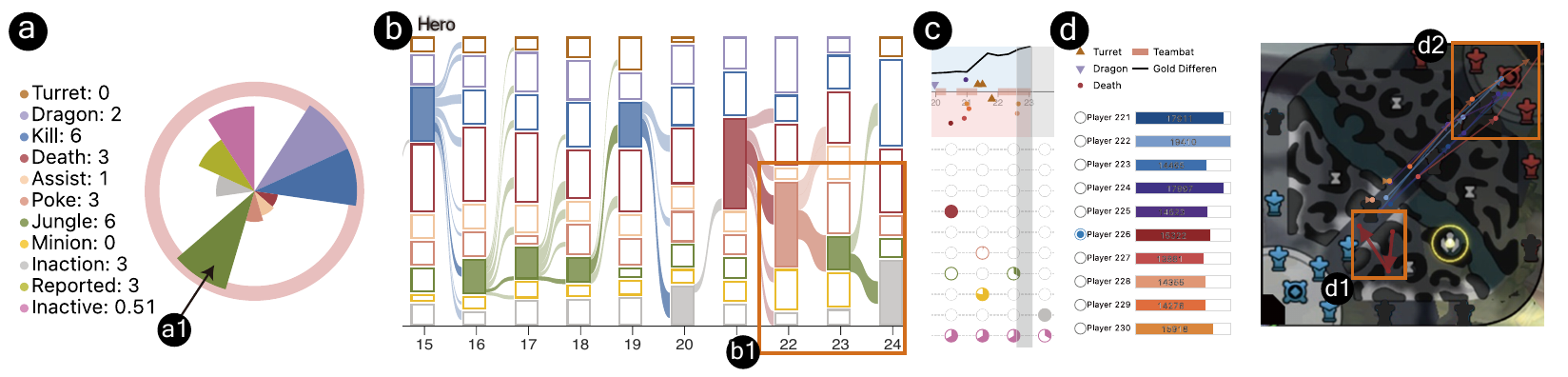}
\caption{This player is identified as an actor with the behavior of not participating in the key defense of the team base. (a) The glyph shows that the player has a large number of monster-killing events. (b) In the hero mode of Progression View, it has been found that the rank of priority events is decreasing. (c, d) A region is brushed, and the map chart demonstrates that the player does not participate in the key defense of the team base in the final stage.}
\label{fig:case_2_3}
\end{figure*}

\textbf{Not participate in the key defense of the team base.} E5 lasso-selected a player with a high number of priority events as killing monster events (\autoref{fig:case_2_3}(a1)) (\textbf{R1}). In the final stage of the game, E5 observed a decrease in the rank of priority events when inspecting the hero mode of {\vtwo} (\autoref{fig:case_2_3}(b1)) (\textbf{R2}).
E5 brushed the region between the $23^{rd}$ and $24^{th}$ minute in the {\vfour} and noticed that while the enemy was advancing towards the team base, the player still focused on killing monsters and failed to participate in the crucial defense of the team base between the $23^{rd}$ and $24^{th}$ minute (\autoref{fig:case_2_3}(c, d1, d2)) (\textbf{R3}).
This negative behavior resulted in the team's inability to successfully defend their base, ultimately leading to the failure of the game. E5 labeled this player as an actor (\textbf{R4}).

\section{Low-level Actor Detection Algorithms}
\label{sec:low_level_actor_detection_algorithms}
\par We introduce two distinct categories of low-level actors in team-based video games: AFK players and feeders. AFK players are those who remain inactive and fail to engage in gameplay, thereby causing an imbalance in the distribution of active players and adversely affecting the overall gaming experience. To identify AFK players, we propose a rule-based algorithm (Algorithm \autoref{alg:AFK}) that accumulates the duration of inactivity in seconds. To account for potential errors in identifying AFK players, expert opinions have been incorporated, and a threshold of 120 seconds has been set to determine whether a player is indeed a low-level actor. On the other hand, feeders deliberately allow their characters to be killed by opposing teams, thereby providing the enemy with an advantage in the game. Algorithm \autoref{alg:feeder} has been developed to detect this behavior, which identifies three specific types of suspected deaths that are correlated with feeding behavior: \textit{turret diving}, \textit{overextending}, and \textit{disguise resistance}. Turret diving refers to a death that occurs when a player sustains all the damage from opposing turrets or is killed by a turret without dealing any damage to opposing heroes. Overextending is defined as a death that occurs when a player receives damage from at least three opponent heroes without offering any resistance. Disguise resistance refers to a death that occurs when actors attempt to conceal their feeding behavior by appearing to resist the opposing team. In such cases, if a player's damage dealt at the time of death is less than 40\% of the damage received, the player is suspected of engaging in feeding behavior. If a player has more than three suspected deaths, they are classified as a low-level actor engaging in feeding behavior.


\begin{algorithm}
\caption{Rule-based AFK detection algorithm.}
\label{alg:AFK}
\begin{algorithmic}
\State \textbf{Input:} $idle \_ time$ \Comment{Duration of time that the player is idle.}
\State \textbf{Output:} $LowLevelActor$ 
\State \textbf{Require:} $threshold = 120$ \Comment{Set the time threshold as 120 seconds.}
\State \textbf{Ensure:} $LowLevelActor \gets false$ \Comment{Initialization}
\If{$idle \_ time \geq threshold$}
    \State $LowLevelActor \gets true$ 
\EndIf
\end{algorithmic}
\end{algorithm}

\begin{algorithm*}
\caption{Rule-based feeder detection algorithm.}
\label{alg:feeder}
\begin{algorithmic}
\State \textbf{Input:} $player\_ death\_ list$ \Comment{Input specific information of the player's deaths.}
\State \textbf{Output:} $LowLevelActor$ 
\State \textbf{Require:} $threshold = 0.4$ \Comment{Set the ratio threshold as 0.4.}
\State $count = 3$ \Comment{Set the count threshold as 3.}
\State \textbf{Ensure:} $LowLevelActor \gets false$ \Comment{Initialization.}
\State $death \_ times \gets 0$ 
\For {$death$ in $player\_ death\_ list$} 
    \State $SuspectedDeath \gets false$
    \If{$player\_ to\_ turret = 0$ and $player\_ to \_hero = 0$ and $hero\_ to\_ player = 0$ and $turret\_ to\_ player \neq 0$} 
    \Comment{Turret diving.}
        \State $SuspectedDeath \gets true$
    \EndIf
    \If{$player\_ to \_hero = 0$ and $hero\_ to\_ player \neq 0$} 
        \If{$dead\_ in \_ turret = true$} \Comment{Turret diving.}
            \State $SuspectedDeath \gets true$
        \ElsIf{$hero\_ number \_ to \_ player \geq 3$} \Comment{Overextending.}
            \State $SuspectedDeath \gets true$
        \EndIf
    \EndIf
    \If{$(player\_ to \_hero + player\_ to \_turret)/(hero\_ to \_ player + turret\_ to \_player)\leq threshold$} 
    \Comment{Disguise resistance.}
        \State $SuspectedDeath \gets true$
        
    \EndIf
    \If{$SuspectedDeath = true$}
        \State $death \_ times \mathrel{+}= 1$
    \EndIf
\EndFor
\If{$death \_ times \geq count$}
    \State $LowLevelActor \gets true$ 
\EndIf
\end{algorithmic}
\end{algorithm*}

\section{Extracted Features for Training XGBoost}
Based on the discussion with the experts, we determined potential features which could be utilized to indicate whether the player is an actor or not. We extracted numerical features for each player to train the XGBoost with the purpose of predicting the label for the remaining unlabeled players. We provided a detailed description of each extracted feature in \autoref{tab:features_list}.

\begin{table*}[]
\centering
\begin{tabular}{ll}
\hline
Feature & Description \\ \hline
gametime & The duration of the match. \\
playerproficiencylv & Player proficiency of the hero. \\
playerherotype & Player hero type. \\
grade & Player rank. \\
roleelo & Player elo score. \\
dmgtotal & Total damage. \\
dmgtohero & Total damage to heroes. \\
towerhurt & Damage to tower. \\
rcvdmgfromall & Received damage from all. \\
rcvdmgfromhero & Received damage from heroes. \\
rcvdmgfromother & Received damage from others. \\
kills & The number of killing heroes. \\
die & The number of deaths. \\
assistant & The number of assisting in killing heroes. \\
coin & Total economy. \\
playermonsterkillcoin & Economy from killing monsters. \\
moneyforkill & Economy from kills or assists. \\
playersoldierkillcoin & Economy from minion killing. \\
killsoldiers & The number of minions killed. \\
battleresult & Battle results. \\
surrendertimes & The number of surrenders. \\
healthyrecall & The number of healthy recall. \\
equiptotalbuy & The number of equipment purchases. \\
playeroffline & The number of offline times. \\
playerreconnection & The number of reconnect times. \\
skillusetimes & The number of skill hits. \\
skillmisstimes & The number of skill misses. \\
playerkilllittledragoncnt & The number of killing dragons. \\
playerkillbigdragoncnt & The number of killing baron. \\
killbluebuff & The number of killing blue buff. \\
killredbuff & The number of killing red buff. \\
triplekill & Triple kills. \\
fourkill & Quadra kills. \\
fivekill & Penta kills. \\
playervisiblewardcount & The number of placing visible wards. \\
idle\_time & The duration of AFK. \\
dmgtohero\_teams\_per & The ratio of total damage to heroes to the team. \\
kills\_teams\_per & The ratio of kills to the team. \\
die\_teams\_per & The ratio of deaths to the team. \\
assistant\_teams\_per & The ratio of assists to the team. \\
coin\_teams\_per & The ratio of the total economy to the team \\
idle\_time\_per & The ratio of idle time to game time. \\
tower\_dead & The number of deaths under the turret. \\ \hline
\end{tabular}
\caption{The description of extracted features used to train XGBoost.}
\label{tab:features_list}
\end{table*}

\end{document}